\documentclass[sigconf,nonacm]{acmart}

\renewcommand\footnotetextcopyrightpermission[1]{} 

\usepackage{float}
\usepackage{subfigure}
\usepackage{listings,amsfonts}
\usepackage{amsmath,pifont}
\usepackage{xspace}
\usepackage{bbding}
\usepackage{balance}
\usepackage{hyperref,endnotes}
\usepackage{booktabs}
\usepackage{array}
\usepackage{algorithm}
\usepackage{multirow,makecell}
\usepackage{enumitem}
\usepackage[normalem]{ulem}
\usepackage{graphicx}
\usepackage{amsthm}
\useunder{\uline}{\ul}{}
\usepackage{multirow}    					
\usepackage{hhline}     				 		

\usepackage{amssymb}
\usepackage{pifont}

\newcommand{\cmark}{\ding{51}}%
\newcommand{\xmark}{\ding{55}}%

\begin{document}

\title{Characterizing Cryptocurrency Exchange Scams}

\author{Pengcheng Xia}
\affiliation{%
  \institution{Beijing University of Posts and Telecommunications, China}
}

\author{Bowen Zhang}
\affiliation{%
  \institution{Beijing University of Posts and Telecommunications, China}
}

\author{Ru Ji}
\affiliation{%
  \institution{Beijing University of Posts and Telecommunications, China}
}

\author{Bingyu Gao}
\affiliation{%
  \institution{Beijing University of Posts and Telecommunications, China}
}

\author{Lei Wu}
\affiliation{%
  \institution{Zhejiang University, China}
}

\author{Xiapu Luo}
\affiliation{%
  \institution{The Hong Kong Polytechnic University}
}

\author{Haoyu Wang*}
\affiliation{%
  \institution{Beijing University of Posts and Telecommunications, China}
}

\author{Guoai Xu}
\affiliation{%
  \institution{Beijing University of Posts and Telecommunications, China}
}

\begin{abstract}
As the indispensable trading platforms of the ecosystem, hundreds of cryptocurrency exchanges are emerging to facilitate the trading of digital assets. 
While, it also attracts the attentions of attackers. 
A number of scam attacks were reported targeting cryptocurrency exchanges, leading to a huge mount of financial loss.
However, no previous work in our research community has systematically studied this problem.
In this paper, we make the first effort to identify and characterize the cryptocurrency exchange scams. 
We first identify over 1,500 scam domains and over 300 fake apps, by collecting existing reports and using typosquatting generation techniques. Then we investigate the relationship between them, and identify 94 scam domain families and 30 fake app families.
We further characterize the impacts of such scams, and reveal that these scams have incurred financial loss of 520k US dollars at least. We further observe that the fake apps have been sneaked to major app markets (including Google Play) to infect unsuspicious users.
Our findings demonstrate the urgency to identify and prevent cryptocurrency exchange scams. To facilitate future research, we have publicly released all the identified scam domains and fake apps to the community.

\end{abstract}

\keywords{Cryptocurrency, Scam, Exchange, Domain Typosquatting, Fake App, Trust-trading}

\maketitle

\section{Introduction}
\label{sec:introduction}

Since the first Bitcoin block was mined back in 2009, cryptocurrency has seen an explosive growth thanks to the evolvement of blockchain technology and their economic ecosystems. Besides BitCoin, thousands of unique cryptocurrencies have popped up from time to time.
As of the end of 2018, there are over 2,000 different cryptocurrencies, and the total market capitalization is \$100bn, which is higher than the GDP of 127 countries~\cite{listofcrypto}. 

As the indispensable trading platforms of the ecosystem, hundreds of cryptocurrency exchanges are emerging to facilitate the trading of digital assets (e.g., Bitcoin) with both traditional fiat currencies (e.g., US dollars) or other digital assets (e.g., Ether).

Inevitably, the prosperity of cryptocurrency exchanges are great targets for hackers to perform attacks to make a profit.
A number of exchanges have been targeted by large-scale hacking attacks. It is reported that the cryptocurrency exchanges suffered a total loss of \$882 million due to targeted attacks in 2017 and in the first three quarters of 2018\cite{attacks2018}. The number keeps increasing in 2019.
For example, as reported in May 2019, attackers have stolen 7,000 bitcoins (which worth \$41m) from Binance, one of the top leading exchanges all over the world, using a variety of techniques, including phishing, viruses and other attacks\cite{binancebreach}. The exchange Coinhouse suffered a phishing attack on September 2019, and attackers gained access to all the user names and email addresses\cite{coinhousephishing}.
It is worth noting that, many attacks are relying on the social engineering 
techniques, i.e., phishing and trust-trading scams.

It is urgent to identify and prevent scam attacks targeting exchanges.
The blockchain community has started to pay attention to the scam attacks in the cryptocurrency ecosystem. For example, several open-source databases (e.g., CryptoScamDB and EtherscamDB) have collected malicious domains and their associated addresses that have the intent of deceiving people for the purposes of financial gain by using a crowd-sourcing based approach (e.g., being actively reported by victims), although only a few of them are related to cryptocurrency exchanges. 
To the best of our knowledge, no previous study in our research community has made efforts to investigate this problem. We are still unaware: 1) to the extent the scams exist in the ecosystem; and 2) who are the attackers behind them; and 3) what are the impacts of the scams.

\textbf{Our Study.} In this paper, we make the first effort to look at the cryptocurrency exchange scams. To cover as much scams as possible, we first use a hybrid approach by first collecting existing known scams and then developing an automated approach, to identify both well-known scams and scams that have not been disclosed to public (see \textbf{Section~\ref{sec:measurement}}). 
We have identified $1,595$ scam domains, and over 60\% of them are not publicly known. 
Besides, we have identified over 300 fake exchange apps. 
Based on the harvested dataset, we propose to cluster the domains and apps, and further investigate the relationship between them (see \textbf{Section~\ref{sec:attackers}}). We have identified 94 scam domain families and 30 fake app families. 
At last, we have investigated the distribution channels of such scams, and their real-world impacts by analyzing their associated blockchain addresses (see \textbf{Section~\ref{sec:impact}}).

In summary, we make the following main research contributions:

\begin{itemize}
    \item \textbf{To the best of our knowledge, this paper is the first systematic study of the cryptocurrency exchange scams}. We collected by far the largest exchange scam dataset, and performed deep analysis of them, including the attackers and impacts. Most of the identified scams have not been known to the community.
    
    \item \textbf{We have revealed that a majority of the scam domains and fake apps were created and controlled by a small number of groups (attackers)}, which could be useful for us to further identify and track the new scams in the future.

    \item \textbf{We have revealed over 182 blockchain addresses related to such scams}. We also identify 518 addresses associated to them, which are quite possible to be controlled by the same group of people. Such information could be used to track the money flow of the scam attacks.
    These scams have incurred financial loss of over 520K US dollars (lower-bound).

\end{itemize}

We have released the scam dataset we collected and all the experiment results to the community at:

\url{https://cryptoexchangescam.github.io/ScamDataset/}

\section{Background and Related Work}
\label{sec:background}

\subsection{Cryptocurrency and Exchange}

Cryptocurrency is a kind of digital asset that uses cryptography to ensure its creation security and transaction security. The first and most well-known cryptocurrency, Bitcoin, was released in 2009, and till now there are over $2,500$ different kinds of cryptocurrencies. With the rise of cryptocurrencies in 2017, people pay more attention on cryptocurrency exchanges in order to get or trade cryptocurrencies. 
A cryptocurrency exchange is a marketplace where users can buy and sell cryptocurrencies. Many of them only offer trade services among cryptocurrencies while a few offer fiat (e.g., US Dollar or Euro) to cryptocurrency transactions. 
Similar to stock market, people flood into cryptocurrency exchanges to invest in order to get the benefit of cryptocurrency price changes. There are three types of cryptocurrency exchenges: centralized exchanges (CEX) which is governed by a company or an organisation, decentralized exchanges (DEX) which provide automated process for peer-to-peer trades, and hybrid exchanges which combine the both of the above.

\subsection{Related Work}

\subsubsection{Blockchain Scams and Attacks}
Blockchain platforms are always the targets of scams and security attacks. 
A few studies have characterized the blockchain scams. Most of them were focused on detecting the ponzi schemes~\cite{chen2018detecting, bartoletti2020dissecting, bartoletti2018data}. Besides, a large number of studies focused on detecting and analyzing attacks from different levels, including blockchain consensus~\cite{bissias2016analysis}, smart contract~\cite{atzei2017survey}, abnormal transactions~\cite{chen2018understanding, EOSIO}, etc. Despite cryptocurrency exchanges are the key infrastructure of the blockchain ecosystem, however, the security-related issues, including the scam problem studied in this work, have not been well-studied yet. 

\subsubsection{Domain Typosquatting}
\label{subsec:typosquatting}
Typosquatting (URL hijacking) is the act of registering a domain name very similar
to an existing legitimate domain, which relies on mistakes such as typos made by Internet users when inputting a website address into a web browser. 
These typosquatting domains are often exploited by attackers. 
Many research studies were focused on detecting and analyzing domain typosquatting. 
Wang et al.~\cite{wang2006strider} proposed a general and widely adopted approach to generate typosquatting domain names. Szurdi et al.~\cite{szurdi2014long} estimated that 20\% of the .com domain registrations are true typo domains and the number is increasing with the expansion of the .com domain space. 
Agten et al.~\cite{agten2015seven} found that even though 95\% of the popular domains we investigated are actively targeted by typosquatting, only few trademark owners protect themselves against this practice by proactively registering their own typosquatting domains.
Besides, a few tools are available to generate possible squatting domains, including URLCrazy~\cite{URLCrazy}, dnstwist~\cite{dnstwist}, etc. 
In this study, we identify that a number of the exchange scams are in the form of typosquatting. Thus, we take advantage of existing techniques to generate typosquatting domains and further analyze the scams (see Section~\ref{subsec:domain}).

\subsubsection{Fake Apps/App Clones}
A \textit{fake} app masquerades as the legitimate one by mimicking the look or functionality. 
Fake apps usually have identical app names or package names to the original ones. 
There have been a number of studies focusing on this topic. 
Wang et al.~\cite{wang2018beyond} proposed a clustering approach on app names to detect potential fake apps. 
Tang et al.~\cite{tang2019large} have characterized over 150K fake apps that have same package names or app names with popular apps. 
Kywe et al.~\cite{kywe2014detecting} and Li et al.~\cite{li2014zoom} proposed technique to detect fake apps based on the external features of apps, e.g., icons, app names.
In this paper, we follow the most traditional method to identify fake apps, i.e., apps share the same app name or app ID (package name) but with different authorship (see Section~\ref{sec:fakeapp}).

\section{Study Design}
\label{sec:studydesign}

In this paper, we perform a large-scale measurement of cryptocurrency exchange scams in the wild. 
We therefore take advantage of various sources and approaches to collect a dataset that covers scams targeting the top cryptocurrency exchanges, in the form of both \textit{domains} and \textit{mobile apps}.

\subsection{Target Cryptocurrency Exchange}
It is first necessary to compile a list of Cryptocurrency Exchanges, which may be subject to scam attacks. As the volume of each cryptocurrency exchange fluctuates greatly every day, the ranking of exchanges is not stable. 
Thus, we resort to Google to first retrieve several ranking lists of Cryptocurrency Exchanges, and then merge them to build a list of 70 popular exchanges, as shown in Table~\ref{tab:overall}\footnote{Due to the ranking fluctuates, we list these exchanges in Table~\ref{tab:overall} based on the dictionary order of exchange names.}. 

To cover both domain and mobile apps, we further collected the official domain names of these exchanges (some exchanges have more than one domain), and their corresponding Android apps\footnote{Note that we only collected their most up-to-date app versions and extract the corresponding developer signatures, which is feasible for us to identify the fake apps in the following study}.

\begin{table}[]
\centering
\caption{The target exchanges and the corresponding results.}
\vspace{-0.1in}
\label{tab:overall}
\resizebox{\linewidth}{!}{

\begin{tabular}{|c|c|c|c|c|c|}
\hline
Name & Launch Time & Official Site & \#Mal URLs & App & \# Fake Apps \\ \hline
Anxpro & Mar-2014 & anxpro.com & 1 & \cmark & 0(0) \\ \hline
B2bx & Oct-2017 & b2bx.exchange & 0 & \xmark & 1(0) \\ \hline
Bcex & Aug-2017 & bcex.ca & 1 & \cmark & 0(0) \\ \hline
Bgogo & May-2018 & bgogo.com & 5 & \cmark & 0(0) \\ \hline
Bibox & Nov-2017 & bibox.com/bibox365.com & 11 & \cmark & 1(1) \\ \hline
Binance & Jul-2017 & binance.com & 320 & \cmark & 16(9) \\ \hline
Bisq & Dec-2014 & bisq.network & 0 & \cmark & 0(0) \\ \hline
Bit-Z & Jun-2016 & bit-z.com & 3 & \cmark & 1(0) \\ \hline
Bitbay & Feb-2014 & bitbay.net & 20 & \cmark & 4(0) \\ \hline
Bitfinex & Oct-2012 & bitfinex.com & 46 & \cmark & 8(4) \\ \hline
bitFlyer & Jan-2014 & bitflyer.com & 13 & \cmark & 0(0) \\ \hline
Bitforex & Jun-2018 & bitforex.com & 7 & \xmark & 1(0) \\ \hline
Bithumb & Jan-2014 & bithumb.com & 5 & \cmark & 5(4) \\ \hline
Bitlish & Jul-2014 & bitlish.com & 1 & \cmark & 0(0) \\ \hline
BitMart & Mar-2018 & bitmart.com & 9 & \cmark & 0(0) \\ \hline
BitMax & Jul-2018 & bitmax.io & 4 & \cmark & 0(0) \\ \hline
BitMEX & Apr-2014 & bitmex.com & 68 & \xmark & 20(8) \\ \hline
Bitpanda & Oct-2014 & bitpanda.com & 44 & \cmark & 2(1) \\ \hline
Bitso & May-2014 & bitso.com & 7 & \cmark & 1(0) \\ \hline
Bitstamp & Jul-2011 & bitstamp.net & 13 & \cmark & 3(1) \\ \hline
Bittrex & Feb-2014 & bittrex.com & 78 & \xmark & 11(5) \\ \hline
BW.COM & Jan-2017 & bw.com & 0 & \cmark & 0(0) \\ \hline
CEX.IO & Jan-2013 & cex.io & 1 & \cmark & 3(0) \\ \hline
Changelly & Oct-2015 & changelly.com & 24 & \cmark & 6(6) \\ \hline
Cobinhood & Dec-2017 & cobinhood.com & 18 & \cmark & 4(0) \\ \hline
CoinAll & Aug-2018 & coinall.com & 0 & \cmark & 0(0) \\ \hline
Coinbase & May-2014 & coinbase.com & 120 & \cmark & 23(15) \\ \hline
Coinbene & Sep-2017 & coinbene.com & 13 & \cmark & 0(0) \\ \hline
Coincheck & Nov-2014 & coincheck.com & 31 & \cmark & 3(3) \\ \hline
Coineal & Apr-2018 & coineal.com & 1 & \cmark & 0(0) \\ \hline
CoinEx & Dec-2017 & coinex.com & 2 & \cmark & 0(0) \\ \hline
CoinExchange & Mar-2016 & coinexchange.io & 0 & \xmark & 12(9) \\ \hline
Coinfloor & Mar-2014 & coinfloor.co.uk & 0 & \xmark & 0(0) \\ \hline
Coinify & Dec-2017 & coinify.com & 0 & \xmark & 0(0) \\ \hline
Coinmama & Apr-2013 & coinmama.com & 25 & \xmark & 7(4) \\ \hline
Coinone & Jun-2014 & coinone.co.kr & 0 & \cmark & 1(0) \\ \hline
Cryptonex & Oct-2017 & cryptonex.org & 3 & \cmark & 0(0) \\ \hline
Cryptopia & May-2014 & cryptopia.co.nz & 3 & \xmark & 23(8) \\ \hline
Deribit & Mar-2015 & deribit.com & 27 & \cmark & 0(0) \\ \hline
DigiFinex & Apr-2018 & digifinex.com & 20 & \cmark & 0(0) \\ \hline
Erisx & Oct-2018 & erisx.com & 9 & \xmark & 0(0) \\ \hline
Etoro & Jun-2011 & etoro.com & 10 & \cmark & 13(2) \\ \hline
EXX & Oct-2017 & exx.com & 1 & \cmark & 0(0) \\ \hline
FatBTC & May-2014 & fatbtc.com & 1 & \cmark & 0(0) \\ \hline
FCoin & May-2018 & fcoin.com & 6 & \cmark & 1(0) \\ \hline
Gate.io & Jan-2013 & gate.io & 1 & \cmark & 2(0) \\ \hline
GBX & Oct-2017 & exchange.gbx.gi & 0 & \cmark & 0(0) \\ \hline
Gemini & Oct-2014 & gemini.com & 10 & \cmark & 0(0) \\ \hline
HitBTC & Dec-2013 & hitbtc.com & 54 & \cmark & 16(7) \\ \hline
Huobi & Sep-2013 & huobi.com/hbg.com & 25 & \cmark & 0(0) \\ \hline
IDAX & Dec-2017 & idax.pro & 1 & \cmark & 0(0) \\ \hline
itBit & Nov-2013 & itbit.com & 6 & \xmark & 0(0) \\ \hline
Kraken & Jul-2011 & kraken.com & 35 & \xmark & 11(2) \\ \hline
KuCoin & Aug-2017 & kucoin.com/kcs.top & 44 & \cmark & 14(7) \\ \hline
LATOKEN & Jul-2017 & latoken.com & 11 & \xmark & 0(0) \\ \hline
Lbank & Oct-2016 & lbank.info & 0 & \cmark & 2(2) \\ \hline
Liquid & Mar-2014 & liquid.com & 0 & \xmark & 0(0) \\ \hline
Livecoin & Mar-2014 & livecoin.net & 11 & \xmark & 2(2) \\ \hline
LocalBitcoins & Jun-2012 & localbitcoins.com & 211 & \xmark & 32(18) \\ \hline
Luno & Feb-2017 & luno.com & 5 & \cmark & 2(2) \\ \hline
OKEx & Jan-2014 & okex.com/okcoin.com & 21 & \cmark & 2(2) \\ \hline
OOOBTC & Nov-2017 & ooobtc.com & 1 & \cmark & 0(0) \\ \hline
Paxful & Jul-2015 & paxful.com & 75 & \cmark & 13(11) \\ \hline
Poloniex & Jan-2014 & poloniex.com & 45 & \cmark & 35(30) \\ \hline
ShapeShift & Jun-2015 & shapeshift.io & 23 & \cmark & 0(0) \\ \hline
Wirex & Dec-2014 & wirexapp.com & 13 & \cmark & 0(0) \\ \hline
Xapo & Mar-2014 & xapo.com & 2 & \cmark & 2(2) \\ \hline
xCoins & Apr-2016 & xcoins.io & 0 & \xmark & 1(1) \\ \hline
Yobit & Aug-2014 & yobit.net & 22 & \xmark & 19(4) \\ \hline
ZB.com & Nov-2017 & zb.com/zbg.com & 8 & \cmark & 0(0) \\ \hline
\end{tabular}
}
\vspace{-0.2in}
\end{table}

\subsection{Research Questions}

Our measurement study in this paper is driven by the following research questions (RQs):

\begin{itemize}
    \item[RQ1] \textbf{Are scam attacks prevalent in the cryptocurrency exchanges?} 
    Although a number of media reports revealed that cryptocurrency exchange scam attacks popped up from time to time, it is still unknown to us to what extent these attacks exist in the ecosystem, and how prevalent are them. 
    Besides, it is also interesting to investigate which cryptocurrency exchanges are their main targets, and how do they perform the scam attacks. 
    We further divide RQ1 into two sub-RQs, RQ1.1: \textit{what is the presence and trend of scam domains?} and RQ1.2: \textit{what is the presence and trend of scam mobile apps?}
    \item[RQ2] \textbf{Who are the attackers behind them?} 
    To understand such attacks in a systematic way, we further want to characterize the real attackers behind them. 
    It is interesting to investigate whether such scams were performed by a group of identical hackers.
    \item[RQ3] \textbf{What is the impact of the scams? }
    Although it is known to us the existence of such scams (e.g., squatting websites and fake apps), it is not clear to us the impact of them, e.g., how many users were tricked and got financial loss.
\end{itemize}

To answer RQ1, we make effort to collect the existing known scams and further identify a large number of unknown scams based on techniques adapted from domain squatting attacks and fake mobile apps.
To answer RQ2, we perform the domain relation analysis based on a set of inherent domain features (e.g., passive DNS, whois, etc.), and the app relation analysis based on the develoepr signature and code-level similarity comparison.
To answer RQ3, we make effort to correlate the scams to blockchain addresses, and collect the transaction information to estimate the number of victims and the amount of financial losses.

\begin{figure*}[t]
\centering
\includegraphics[scale=0.5]{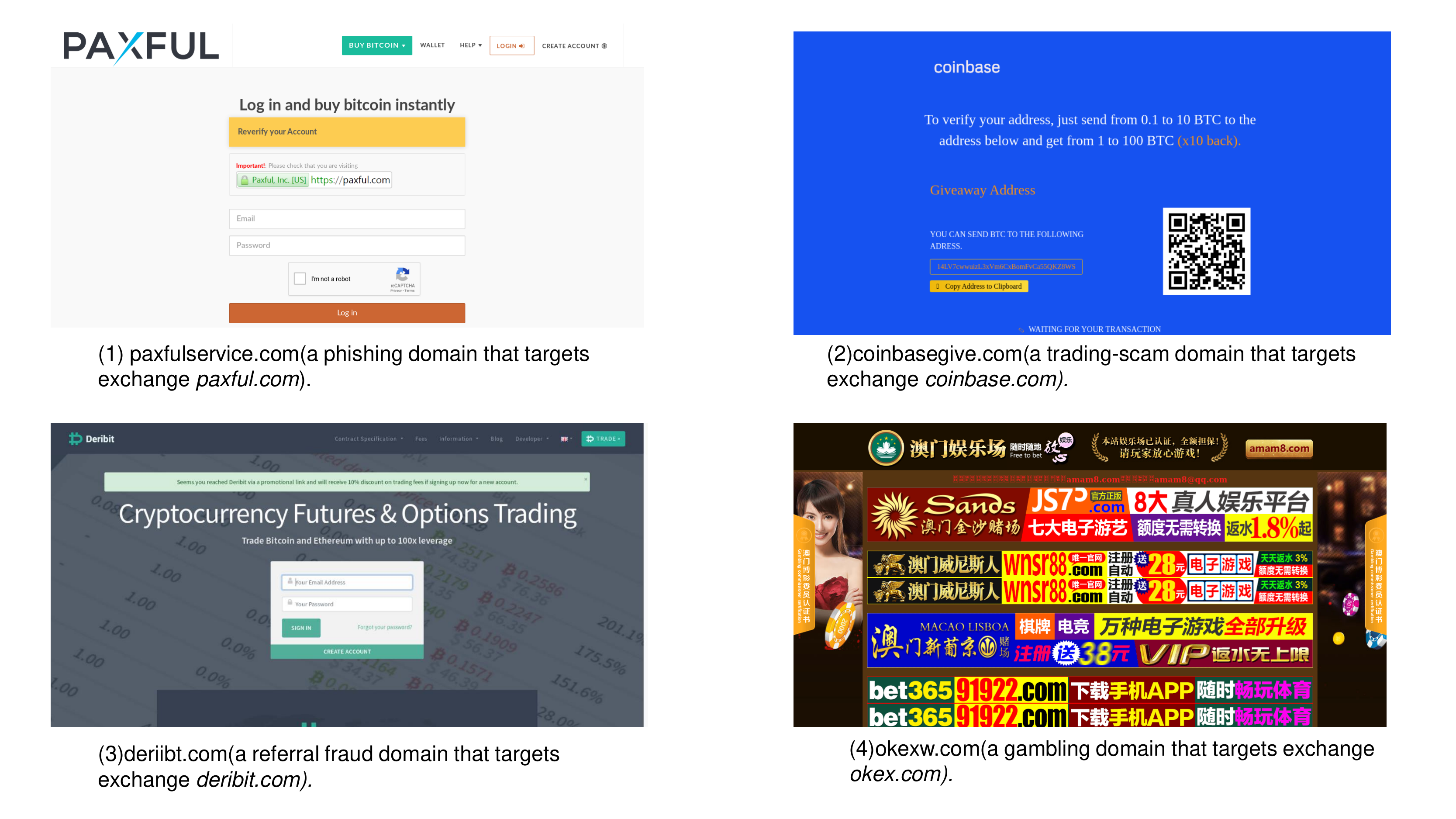}
\caption{Examples of Scam Domains.}
\label{fig:domainmotivating}
\end{figure*}

\section{Measurement of the Scams}
\label{sec:measurement}

In this section, we measure the presence of exchange scams in the form of both domain scams and app scams. To cover as much scams as possible, we use a hybrid approach here, by collecting the existing known scams first and then develop automated approaches to further identify scams that have not been disclosed to public.

\subsection{Detecting the Scam Domains}
\label{subsec:domain}

\subsubsection{Collecting Scam Domains from Existing Corpus}
There are some well-known websites collecting scam cryptocurrency domains in our community, e.g., \textit{etherscamdb.info} and \textit{cryptoscamdb.org} are two representative ones.
Thus, we first write crawlers to collect the known scam domains, and then filter exchange related ones. To this end, 657 scam exchange domains were collected using this approach by the time of our study.

\subsubsection{Generating the Squatting Domains}
By manually exploring the collected scam domains, we have identified that a number of them were distributed using the domain typosquatting techniques. 
They are mainly using these domains to create websites looking exactly similar with the correct one, resulting in the loss of users' credentials or assets.

Thus, we further explore whether there are more scam domains that have not been disclosed to public. As domain squatting has been widely studied in our community and there are many tools available, we take advantage of \textit{dnstwist}\cite{dnstwist}, a widely used tool to generate typosquatting domains and identify the scam ones~\cite{dnstwist1, dnstwist2}. Dnstwist has embedded 13 generation models to explore the possible squatting domains. Take domain \textit{binance.com} as an example, over 2,000 possible squatting domains would be generated using different transformation methods, such as \textit{addition} (e.g., \textit{binancer.com}), bitsquatting (e.g., \textit{bijance.com}), \textit{homoglyph} (e.g., \textit{bin$\alpha$nce.com}), \textit{hyphenation} (e.g., \textit{bi-nance.com}), \textit{insertion} (e.g., \textit{bibnance.com}), \textit{omission} (e.g., \textit{binace.com}), \textit{repetition} (e.g., \textit{binancce.com}), \textit{replacement} (e.g., \textit{binancw.com}), \textit{subdomain} (e.g., \textit{binan.ce.com}), \textit{transposition} (e.g., \textit{binanec.com}), \textit{vowel-swap} (e.g., \textit{binonce.com}), \textit{various} (e.g., \textit{binancecom.com}) and \textit{dictionary} (e.g., \textit{my-binance.com,binancepay.com}).

In this way, we feed the domains of the 70 studied exchanges to \textit{dnstwist}, and we have generated 144,392 squatting candidates in total. Note that, as some domains have not been registered, thus we further filter the domains that have no corresponding IP addresses during our visiting. 
At last, we have identified $4,457$ valid domains by the time of our study (2019-09-23).

\subsubsection{Labelling the Domains}
\label{subsec:domainlabel}
Note that, not all the squatting domains deliver the malicious or scam purposes, as some of them are only used for parking services~\cite{alrwais2014understanding, metcalf2014domain}. 
Thus, we further seek to label the suspicious domains and identify the malicious ones. 
We collect all the possible information related each domain, including the \emph{WHOIS} information, \emph{DNS} information, \emph{autonomous system numbers} and \emph{VirusTotal anti-virus engine scan results}\footnote{VirusTotal (https://www.virustotal.com/) is a widely-used online anti-virus service that combining over 60 state-of-the-art engines.}. 
Furthermore, we write crawler to get the \emph{screenshots} of these websites, the \emph{source code} of webpage, and record the \emph{redirect links}.
Then, we follow the most widely used approach~\cite{quinkert2019s} in our community, to label the domains in an semi-automated way. 

First, as some domains display only blank pages during our visiting, thus we remove such domains (labelled as \textbf{C1: Registered}). Then, for each domain, we analyze the landing URL (the page that one URL is finally redirected to), source code and screenshots, by comparing them with the ones of known parking services and their corresponding official websites, to determine whether they are using parking service or redirect users to their referral links (labelled as \textbf{C2: Parked} and \textbf{C5: Referral Fraud}).
After that, we take advantage of OCR techniques to analyze the content similarity and image similarity, between these websites and their corresponding official websites, in order to identify the phishing websites(labelled as \textbf{C3: Phishing}).
We also rely on VirusTotal's labelling results to classify if a domain is used for phishing and scamming purposes. For the domains flagged by VirusTotal, we further manually analyze them to classify them into phishing (labelled as \textbf{C3: Phishing}) or trading scam (labelled as \textbf{C4: Trading Scam}).
Furthermore, we collect all the image contents listed on the domains to identify whether they are used to perform devious behaviors (e.g., adult and gambling) using Google Cloud Natural Language API and Vision API\footnote{Google Cloud APIs (https://cloud.google.com/apis/) provide Natural Language API to help understand the texts and Vision API to help identify explicit images} (labelled as \textbf{C6: Adult and Gambling}).
At last, for the remaining unclassified domains, we perform manually analysis to see whether they belong to the aforementioned categories or not. Note that, some of the generated domains may be false positives, i.e., they are benign and their names are authentic, which will be flagged during the manually verification (labelled as \textbf{C0: False Positive}).

In this way, we are able to classify the scam domains into the following categories:

\begin{itemize}
    \item \textbf{C0 False Positive}:
    There are 728 domains (14.30\%) belonging to this category, which were flagged during manually verification. Their names are authentic and they are benign websites. 
    For example, the domain name \textit{https://bid\-flyer.com/} looks like bitFlyer's domain, while it is an auction platform for airlines.

    \item \textbf{C1 Registered}: 
    Although some domains have corresponding resolved IP addresses, while they cannot reached during our experiments or they just display blank pages. 
    Thus, we label such domains as 'Registered'.
    Roughly 23.54\% (1,198) domains belong to this category.
    
    \item \textbf{C2 Parked}: The domains using parking services account for 30.83\% of our dataset. People who hold domains usually use parking services to advertise or sale their domains.
    
    \item \textbf{C3 Phishing}: Phishing domains account for 8.35\% of our generated dataset. They often have similar looks with the official ones, making it easier for users to be tricked into typing in their account credentials or downloading malware the domains provide. In our dataset, Binance exchange has the most number of phishing domains (107 domains).
    
    \item \textbf{C4 Trading Scam}: 
    These domains tend to directly take users' money or digital assets. Among 249 Trading scam cases, 232 domains are the Trust-Trading scams. A trust-trading occurs when a victim gives a scammer money (e.g., BTC or ETH), trusting that the scammer will then return them with high-level interest rate investment or rich payback. Instead, however, the attackers simply take the victim's money and leave. Other cases of this category include offering fake exchange support channel or Ponzi schemes, etc.  
    
    \item \textbf{C5 Referral Fraud}: The Referral Fraud domains account for 16.42\% of our dataset. This kind of domains often forwards users to the official exchanges' domain while adding attackers' affiliate code in order to earn a reward provided by these exchanges' referral program.
    
    \item \textbf{C6 Adult and Gambling}: We find 85 domains redirect users to adult or gambling websites. Although these websites have almost no relations with crptocurrency exchanges, they create these typosquatting domains with the malicious purpose of attracting users.
\end{itemize}

In this paper, we regard the last four categories (phishing, trading scam, referral fraud, adult and gambling) as scam domains in general, as all of them fulfill either scam or malicious purposes.
Figure~\ref{fig:domainmotivating} shows the four representative examples of scam domains.

\subsubsection{Overall Result}
At last, we have identified 1,595 scam domains, and 58 exchanges (83\%) were targeted by them. 
Note that only 657 domains have been reported on existing scam databases, and \emph{over 58.8\% of them have not been disclosed to our community}.

\begin{figure}[t]
\centering
\includegraphics[width=0.45\textwidth]{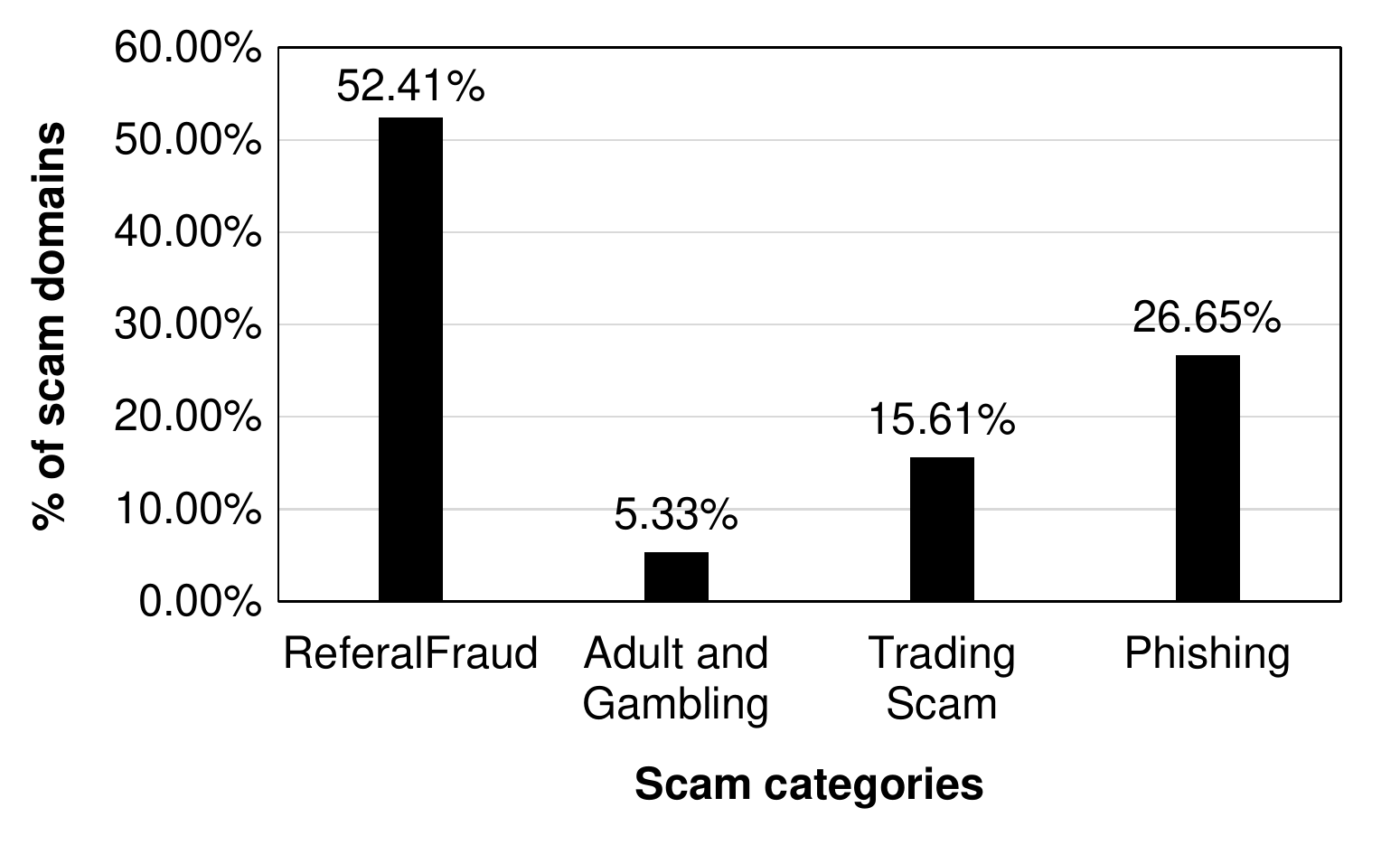}
\caption{The distribution of scam domains.}
\label{fig:domaindistribution}
\end{figure}

\textbf{General Distribution.} 
The distribution of these domains is shown in Figure~\ref{fig:domaindistribution}. Referral Fraud is the most popular category, representing 52.41\% of all the scam domains. Phishing is the second largest category, targeting 28 exchanges. Besides, we have identified 249 trading scam domains, targeting 21 exchanges.

\begin{figure}[t]
\centering
\includegraphics[width=0.45\textwidth]{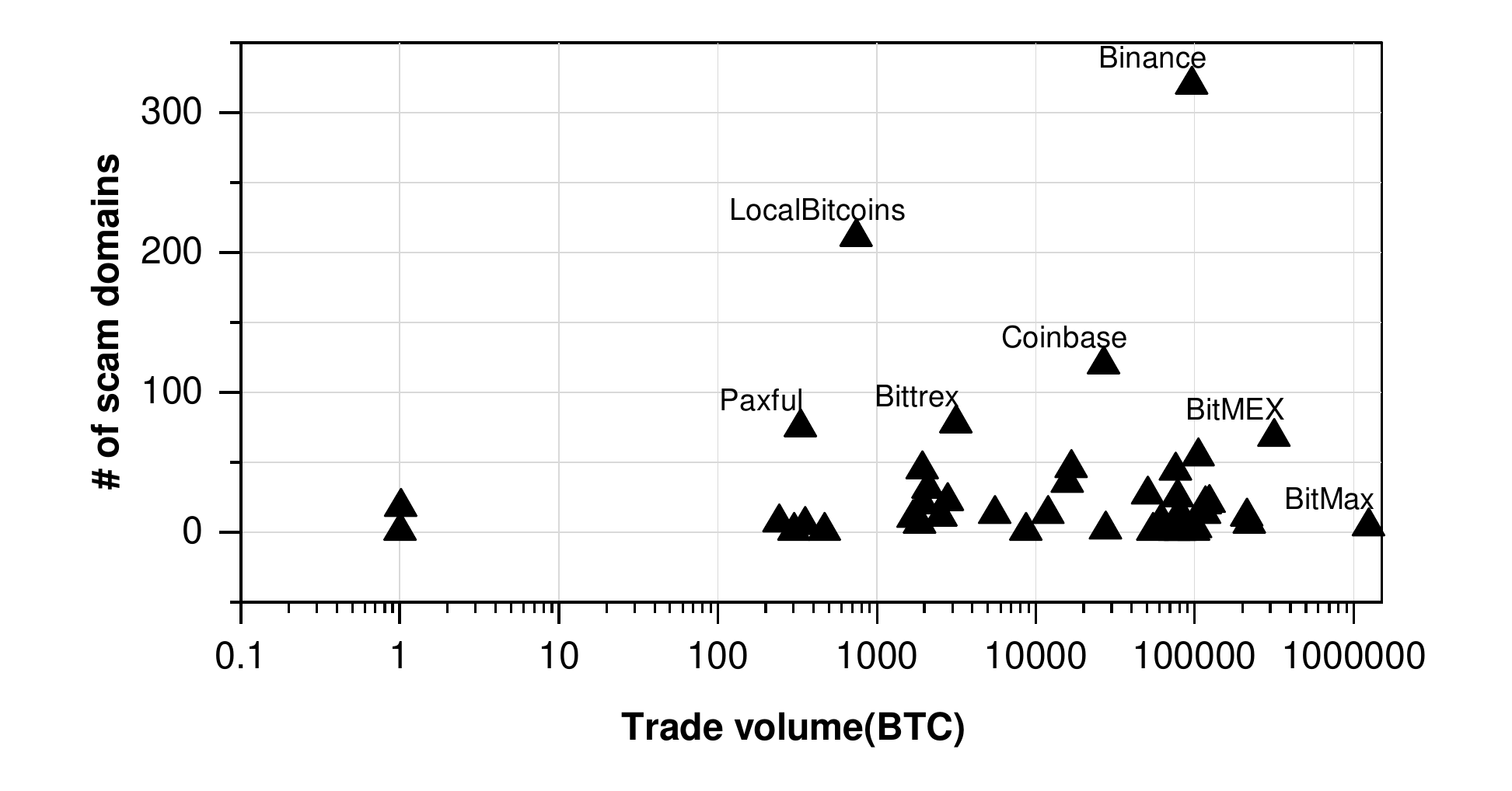}
\caption{The volume of Exchange VS. the number of scam domains.}
\label{fig:targetexchange}
\end{figure}

\begin{figure}[t]
\centering
\includegraphics[width=0.45\textwidth]{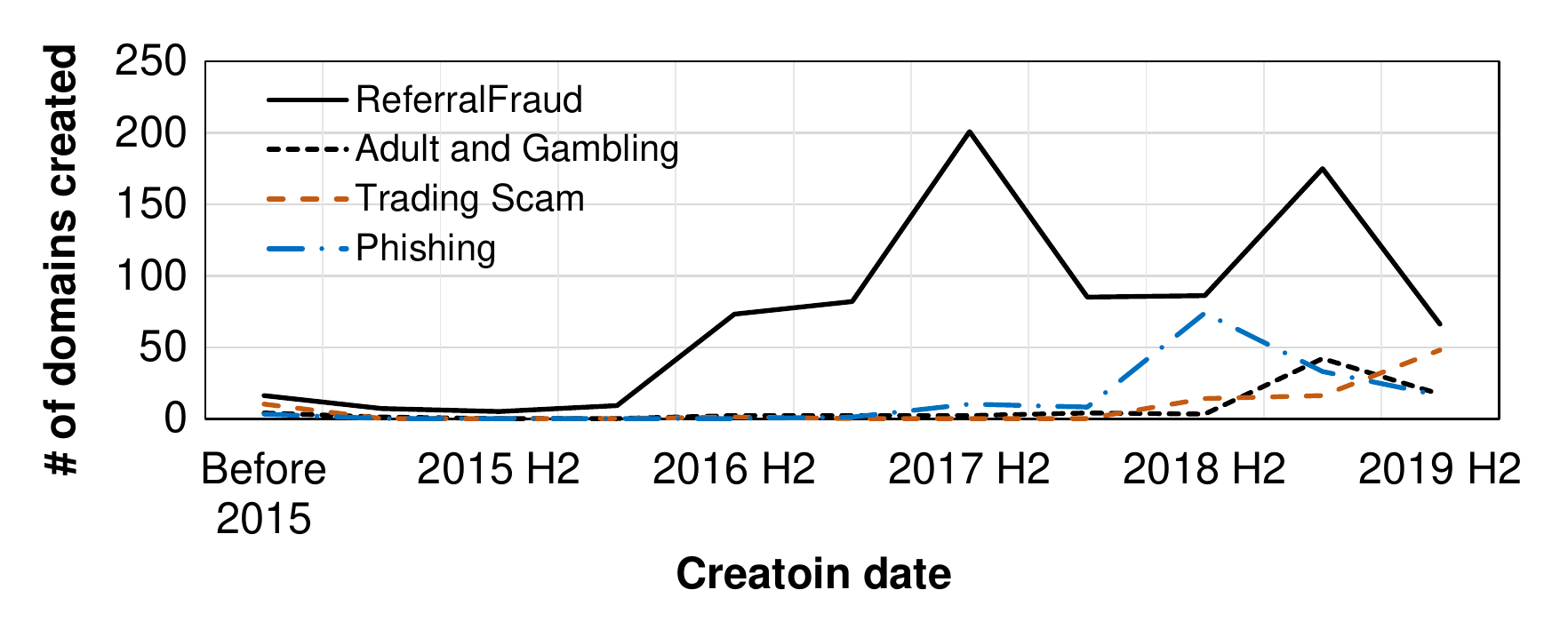}
\caption{The distribution of creation time of scam domains.}
\label{fig:evolution}
\end{figure}

\textbf{Target Exchanges}
Figure~\ref{fig:targetexchange} shows the distribution of targeted exchanges, and the relation with the exchanges' volume\footnote{As the trading volume of exchanges change everyday, here we use the volume of 2019-08-01 to represent each exchange.}. 
Binance, LocalBitcoins, and Coinbase are the exchanges that have the most number of scam domains. It is interesting to see that, in general, scam domains mainly target the exchanges with high volume, while not all popular exchanges have a large number of scam domains. For example, BitMax has the largest trading volume on 2019-08-01, while it only has 4 scam domains.
The reason might be that, BitMax becomes popular since mid 2019, and it has released an incentive plan to attract users\footnote{https://www.cryptoglobe.com/latest/2019/09/one-exchange-racked-up-50-of-all-tether-volume-in-august/}. Before that, it is not as popular as other major exchanges. Thus, we find only a few scams of it.

\textbf{The Evolution of Scam Domain.}
\label{sec:urlevolution}
We further analyzed the evolution of scam domains, as shown in Figure~\ref{fig:evolution}. 
We use the creation date of WHOIS information to represent when a domain was appeared. 
As expected, the number of scam domains has increased rapidly after 2017, following the explosive growth of blockchain techniques. 
It is surprising to observe that, the first exchange scam domain was found in 2004-04-08. 
However, there was no exchange by the time of 2004. 
Our manually verification suggests that this is not a false positive.
The domain name is \textit{www.etorro.com}, which is verified to be a Referral Fraud domain targeting Etoro exchange. 
Thus, we guess that the domain turns to be the referral fraud after Etoro was founded in 2011), before that it might be a domain with other purposes.

\begin{figure}[htbp]
\centering
\includegraphics[width=0.45\textwidth]{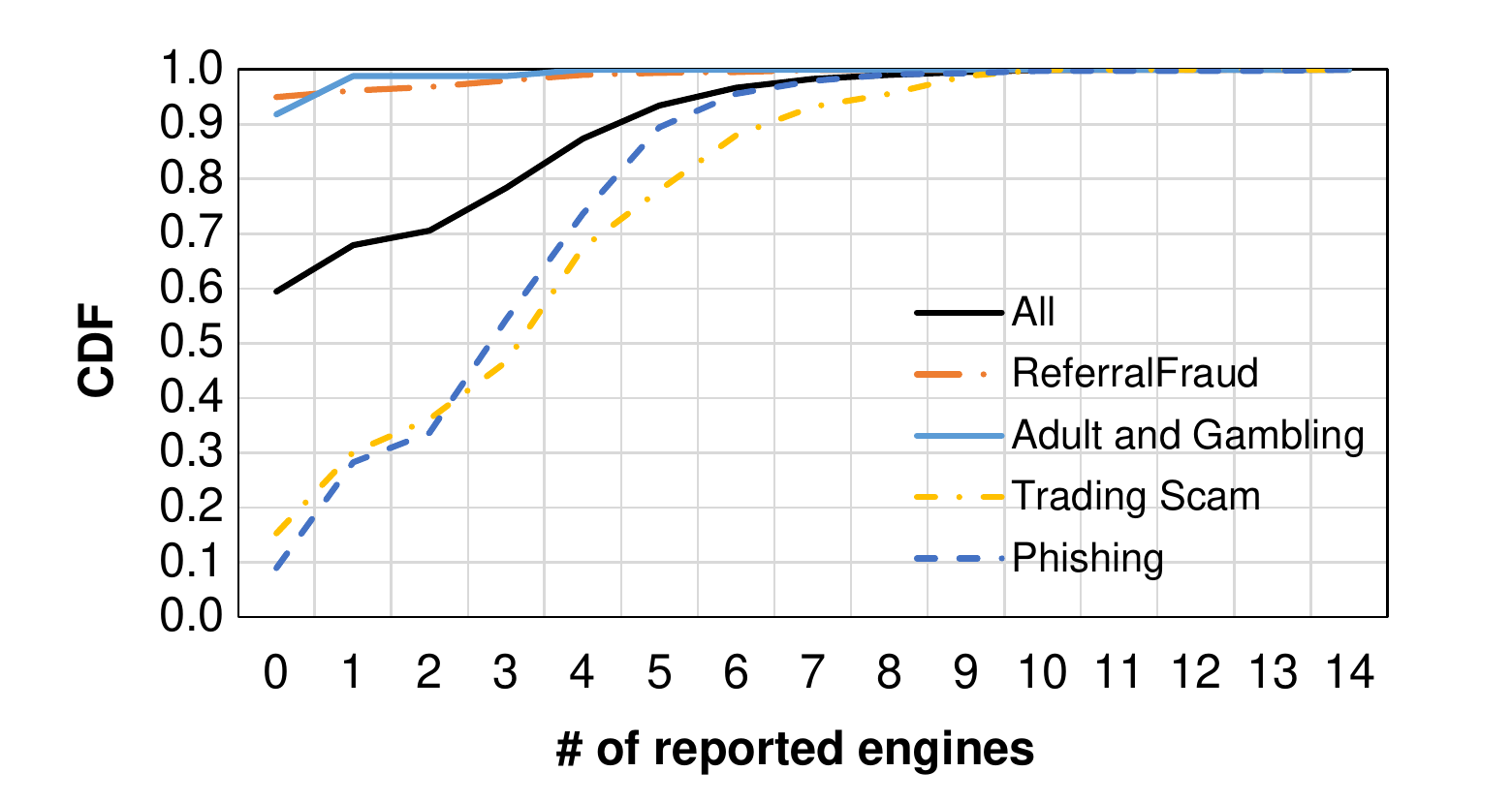}
\caption{The distribution of the number of flagged VT engines for our collected scam domains.}
\label{fig:VTDomain}
\end{figure}

\textbf{How many of them are flagged by anti-virus engines?}
As shown in Figure~\ref{fig:VTDomain}, it is surprising to see that, over 60\% of the domains in our dataset have not been flagged by any anti-virus engine on VirusTotal and only 40.56\% of the domains are flagged by at least 1 engine.
As for each category, over 90\% of ReferralFraud domains and 90\% of Adult and Gambling domains are not detected by anti-virus engines.
Although most of the Trading Scam and Phishing domains are labelled by at least 1 engines, only very few of them are labelled by 10 or more engines\footnote{As previous work suggested that some anti-virus engines on ViusTotal may not always report reliable results, thus they tend to choose a threshold (e.g., 5 or 10).}.
Table~\ref{tab:urlVTtop5} shows the top-5 domains ranked by the most number of flagged engines.

\begin{table}[t]
\caption{Top-5 domains ranked by the most number of enginges on VirusTotal.}
\resizebox{\linewidth}{!}{

\begin{tabular}{@{}cccc@{}}
\toprule
Domain & Target exchange & \begin{tabular}[c]{@{}c@{}}\# engines \\ reported\end{tabular} & Category \\ \midrule
xn--localitcoins-bh4f.net & LocalBitcoins & 14 & Phishing \\
paxfuyl.com & paxful & 11 & ReferralFraud \\
yobit.tilda.ws & yobit & 10 & Trading Scam \\
binance.eth-win.com & Binance & 10 & Trading Scam \\
binancepromo-now.online & Binance & 10 & Trading Scam \\ \bottomrule
\end{tabular}
}
\label{tab:urlVTtop5}
\end{table}

\vspace{0.1in}
\noindent\fbox{
	\parbox{0.95\linewidth}{
		\textbf{Answer to RQ1.1:} 
Our experiment results suggested that the scam exchange domains are prevalent in the ecosystem. Over 83\% (58) of our studied exchanges are targeted by 1,595 scam domains, and most of them were used for malicious purposes including phishing, trading scam, referral fraud, adult and gambling. We have identified 938 domains that have not been disclosed to our community. Unfortunately, most of the domains cannot be flagged by existing anti-virus engines on VirusTotal.
}}

\subsection{Detecting the Fake Apps}
\label{sec:fakeapp}

\subsubsection{Identifying Fake Apps}
To identify fake exchange apps, we first make efforts to collect all the most up-to-data apps from the official websites for each exchange, and extract the certificate signatures from apps\footnote{We assume that each developer will use the same signature to sign their apps and developers' privacy keys will not be leaked, which is a commonsense adopted by the research community.}. 
Then we seek to search all the possible fake apps from app markets. Note that, as app market such as Google Play always removes malicious and fake apps from time to time, it is hard for us to compile a complete list of fake exchange apps. Here, we resort to Koodous\footnote{https://koodous.com/}, a large Android app repository with over 53 million apps in total by the time of our study, containing apps from various sources including Google Play. We use crawler to search the app names and package names in Koodous, and collect all the related apps with same app names or package names. For the collected apps, we further analyze their developer signatures and compare them with the original ones. If found mismatch, we then regard them as fake apps. Note that, this is the general method used in our community to identify fake apps.

\subsubsection{Overall Result}

We have collected 2,810 apks from Koodous, and 323 of them are fake apps -- have same app name/package name with the official exchange apps but signed by different developer signatures. The other apps are official apps with different versions released by the exchanges.

\begin{figure}[t]
\centering
\includegraphics[width=0.45\textwidth]{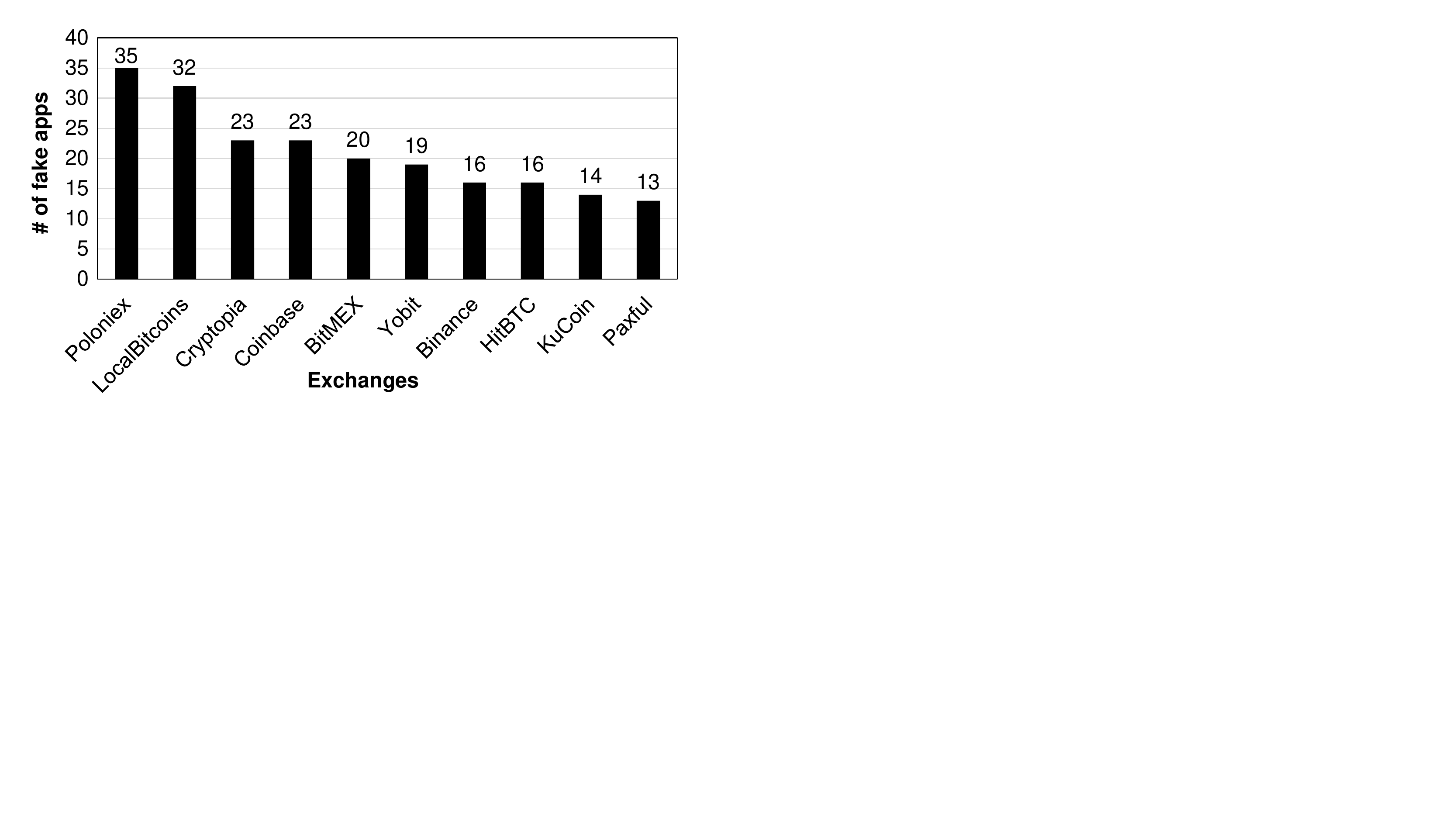}
\caption{Top 10 targeted exchanges of Fake Apps.}
\label{fig:topfakeapp}
\end{figure}

\begin{figure}[htbp]
\centering
\includegraphics[width=0.45\textwidth]{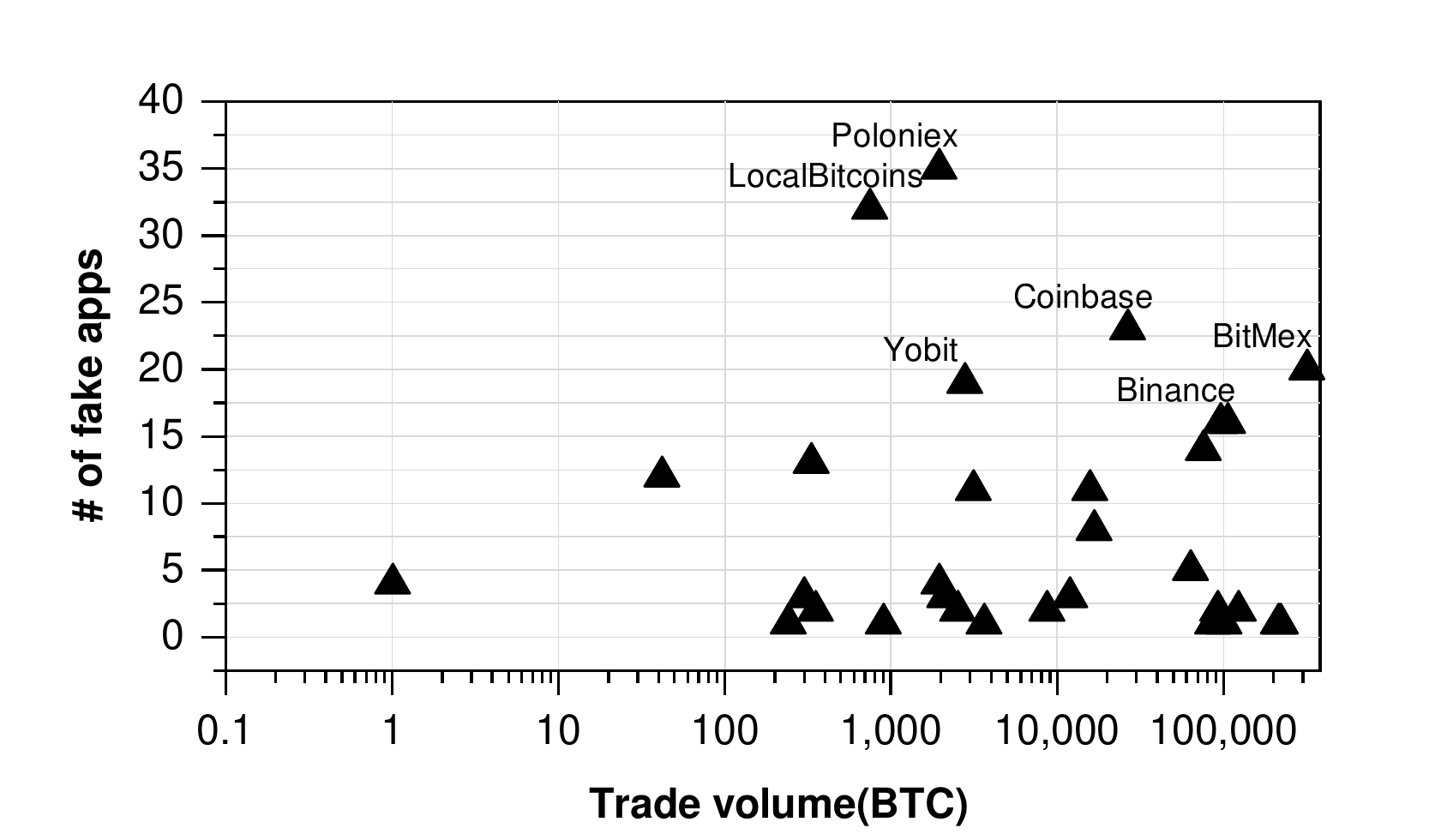}
\caption{The volume of exchange vs. the number of fake apps.}
\label{fig:appdistribution}
\end{figure}

\textbf{Distribution of Fake apps.} 
These fake apps target 38 exchanges (54\%) in total. Figure~\ref{fig:topfakeapp} shows the top-10 targeted exchanges. 
Poloniex exchange has 35 fake apps in total (with 45 scam domains as shown in Figure~\ref{fig:domaindistribution}). 
For the top-10 targets of fake apps, 7 of them are the same with that of scam domains.
We further investigate whether the popular exchanges would receive more fake apps. 
As shown in Figure~\ref{fig:appdistribution}, the general trend is similar with that of scam domains, i.e., fake apps usually target popular exchanges with large trading volume. But there are exceptional cases too.
For example, BitMax has no corresponding fake apps in our study, while it has only 4 scam domains. As aforementioned, the reason might be that BitMax becomes popular since mid 2019 due to its incentive mechanism introduced, and it has not received much attentions from attackers by the time of our study.

\begin{figure}[t]
\centering
\includegraphics[width=0.45\textwidth]{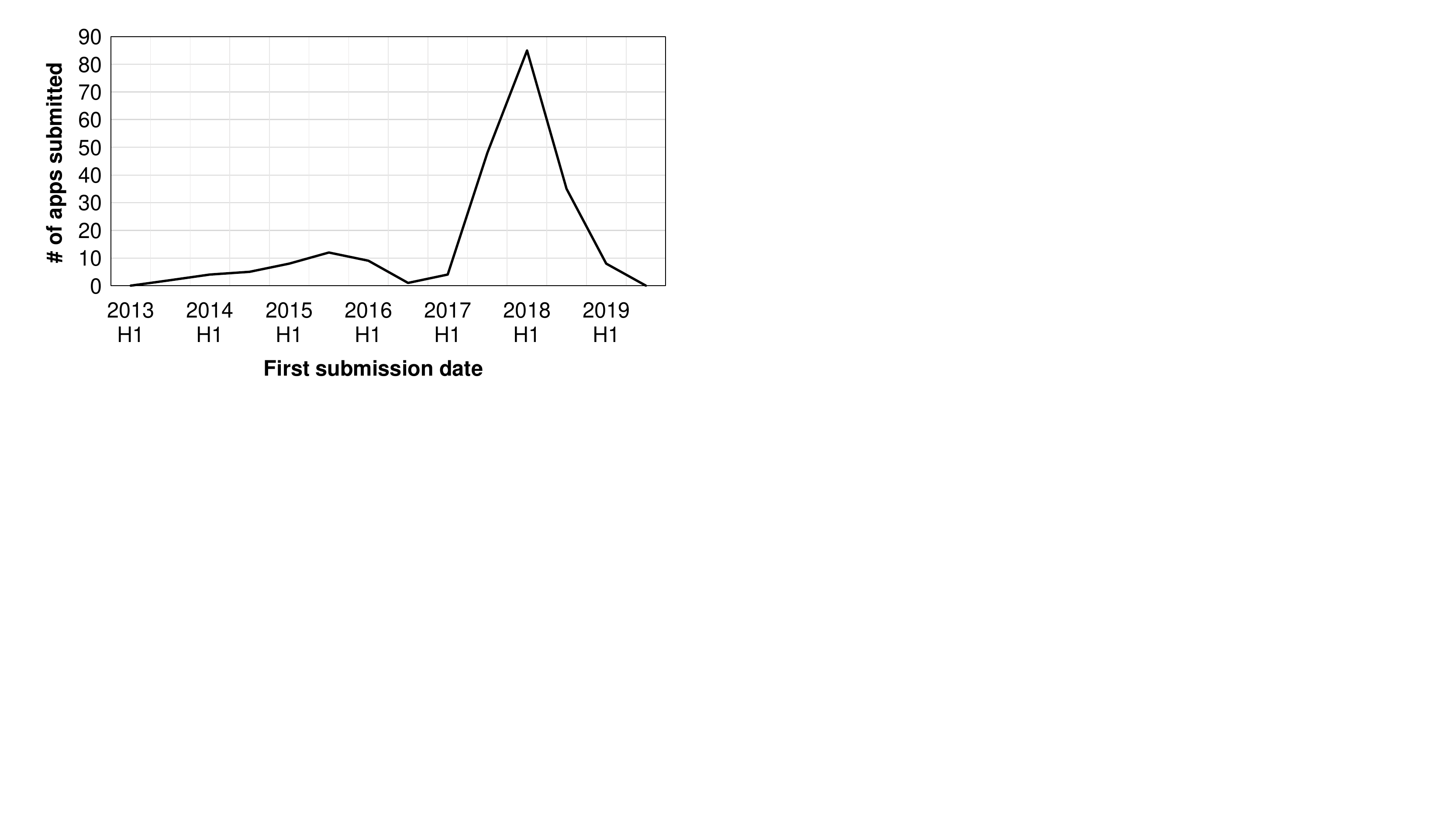}
\caption{The evolution of fake apps.}
\label{fig:appevolution}
\end{figure}

\begin{figure}[t]
\centering
\includegraphics[width=0.45\textwidth]{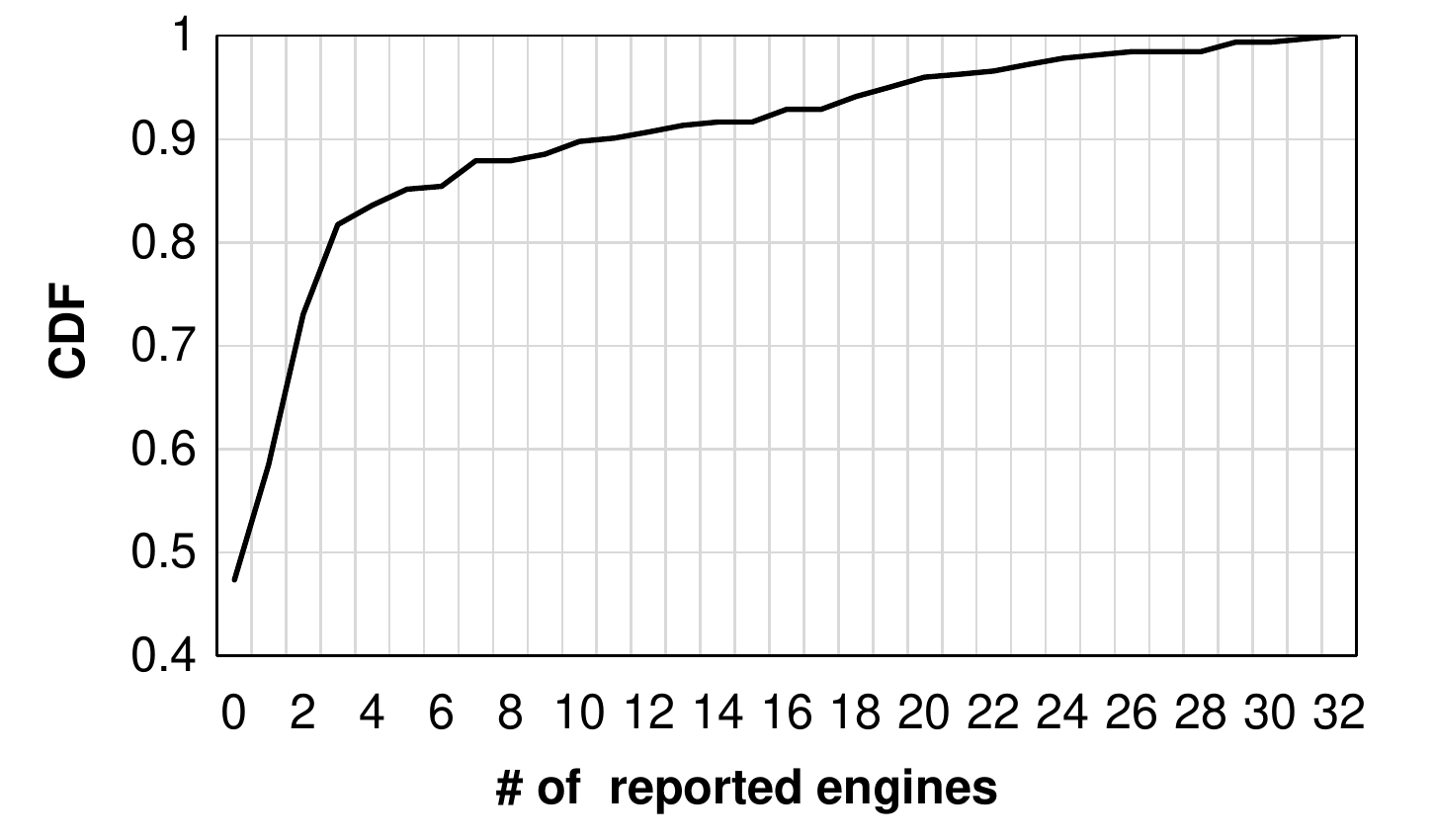}
\caption{\# of VirusTotal reported engines when scanning fake apps.}
\label{fig:appVT}
\end{figure}

\textbf{The Evolution of Scam Apps.}
We refer to the \textit{first seen} time on VirusTotal to show the evolution of scam apps, and the distribution is shown in Figure~\ref{fig:appevolution}. 
The first fake exchange app in our dataset appeared on Nov 16th 2013, targeting at \textit{Etoro exchange} with a referral link. In our dataset, fake apps began to appear in the second half of 2013 and reached its peak of 85 in the first half of 2018.

\textbf{How many of them are flagged by anti-virus engines?}
We further analyze how many of the fake apps are flagged by anti-virus engines. As shown in Figure~\ref{fig:appVT}, over 52.6\% (170) of them are flagged at least one engine, and 33 apps are flagged by over 10 engines. Table~\ref{tab:appVTtop5} shows top-5 of them.

\begin{table}[t]
\caption{Top-5 fake apps ranked by the number of flagged engines on VirusTotal.}
\resizebox{\linewidth}{!}{

\begin{tabular}{@{}cccc@{}}
\toprule
App name & \begin{tabular}[c]{@{}c@{}}Target \\ exchange\end{tabular} & md5 & \begin{tabular}[c]{@{}c@{}}\# engines\\ reported \end{tabular}\\ \midrule
Bitcoin allet - Coinbase & Coinbase & d41d8cd98f00b204e9800998ecf8427e & 32 \\
Binance Secured & Binance & 487ad3a4d18c8b2274bff5916c67bee9 & 31 \\
Bithumb update & Bithumb & e7f634c53f0f0ddd48503d4efb661824 & 29 \\
Bitcoin allet - Coinbase & Coinbase & 76c691abacd276642f11041ec2f78355 & 29 \\
Coinbase & Coinbase & b9f6d2c42e961330dfed437f068a6bb1 & 29 \\ \bottomrule
\end{tabular}
}
\label{tab:appVTtop5}
\end{table}

\subsubsection{Classification of Fake apps.}

\begin{figure}[t]
\centering
\includegraphics[width=0.45\textwidth]{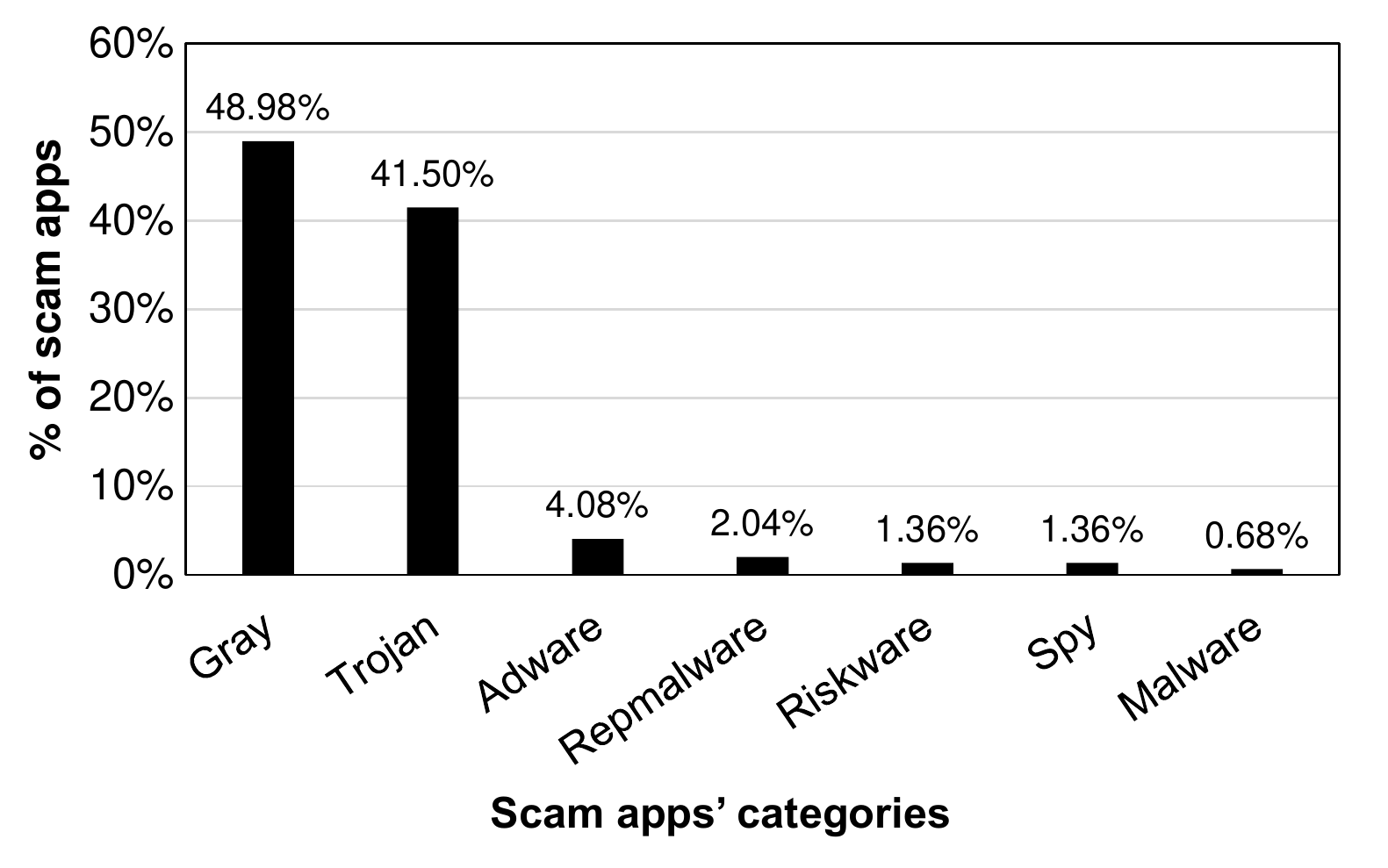}
\caption{The distribution of malware types.}
\label{fig:apptype}
\end{figure}

\begin{figure}[t]
\centering
\includegraphics[width=0.45\textwidth]{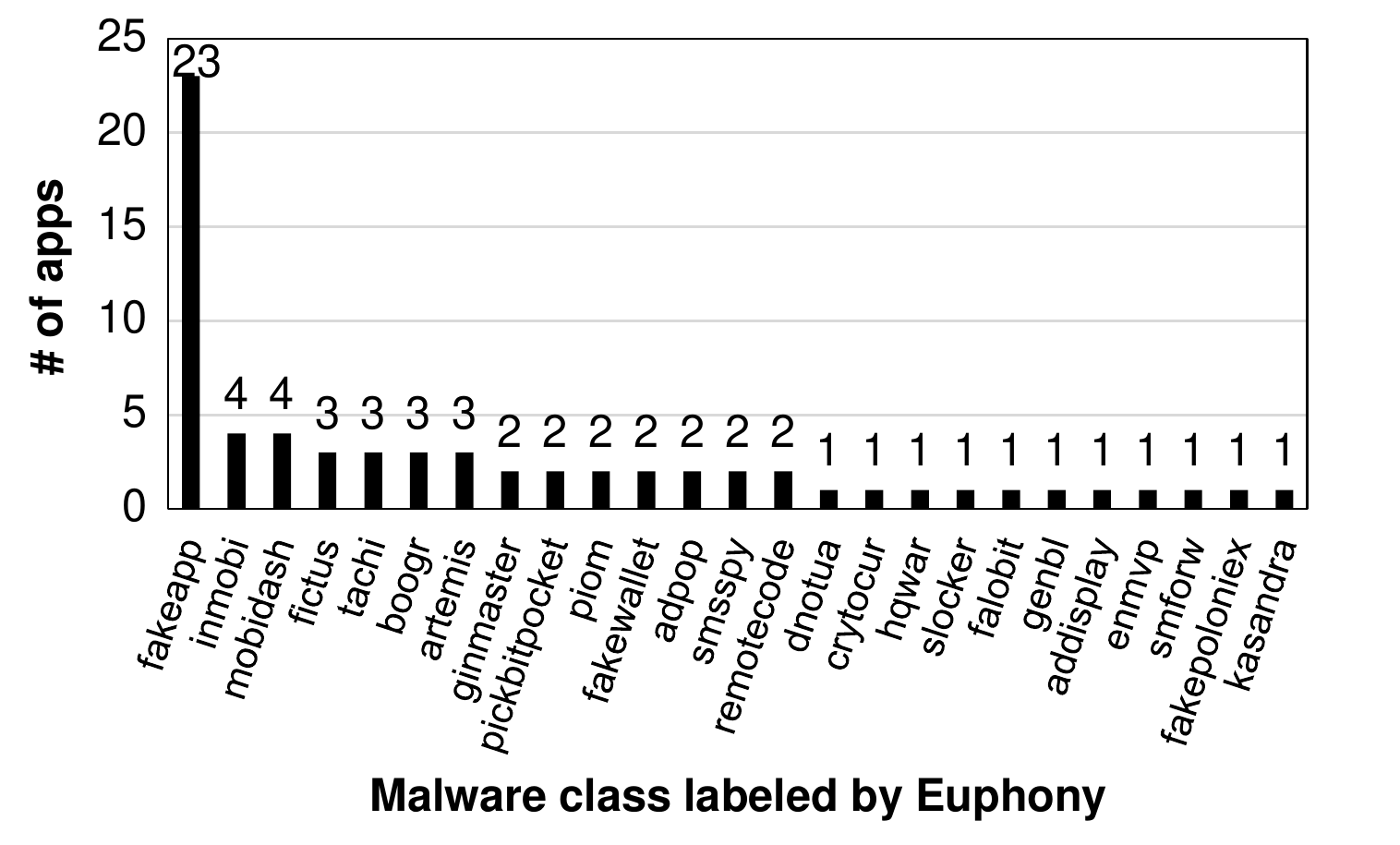}
\caption{The distribution of malware families.}
\label{fig:malwarefamily}
\end{figure}

To classify the fake apps, we use two complementary approaches.
For the 170 apps that were flagged by VirusTotal, we use Euphony~\cite{hurier2017euphony}, a widely-used tool to analyze the scanned results to label a malware type and malware family for each of them. 
For the remaining 153 apps that were not flagged by VirusTotal, we either install them on smartphones or decompile them using static analysis tools for manually examination.

\textbf{Malware Type and Malware Family Distribution.}
As shown in Figure~\ref{fig:apptype}, for the 170 flagged fake apps, roughly 50\% of them are labelled as \textit{grayware} by VirusTotal. Over 40\% of them are flagged as \textit{Trojan}, and roughly 4\% of them are labelled as \textit{adware}. This result suggests that \textit{these fake apps may expose great security threats to users}.
To be specific, we use Euphony to generate a malware family label for each of them, and Figure~\ref{fig:malwarefamily} shows the malware family distribution. As expected, family \emph{fakeapp} ranks the first, with 23 apps labelled. The labels of the remaining apps vary greatly, including adware families like \textit{inmobi}, \textit{mobidash}, and malicious categories like \textit{smsmspy} and \textit{slocker}.
Note that, there are 102 apps that Euphony cannot give them a family label based on the flagged results of engines. This result also suggests that existing anti-virus engines cannot classify these fake apps accurately.

\textbf{Manually Inspection.}
For the remaining 153 fake apps, we found that all of them are referral apps with advertisements. 
Similar with referral fraud domains, most of the referral apps use webview to connect to their referral links, intending to attract new users.
In general, they will also embed some ad libraries to increase the income.

\begin{figure}[t]
\centering
\includegraphics[width=0.45\textwidth]{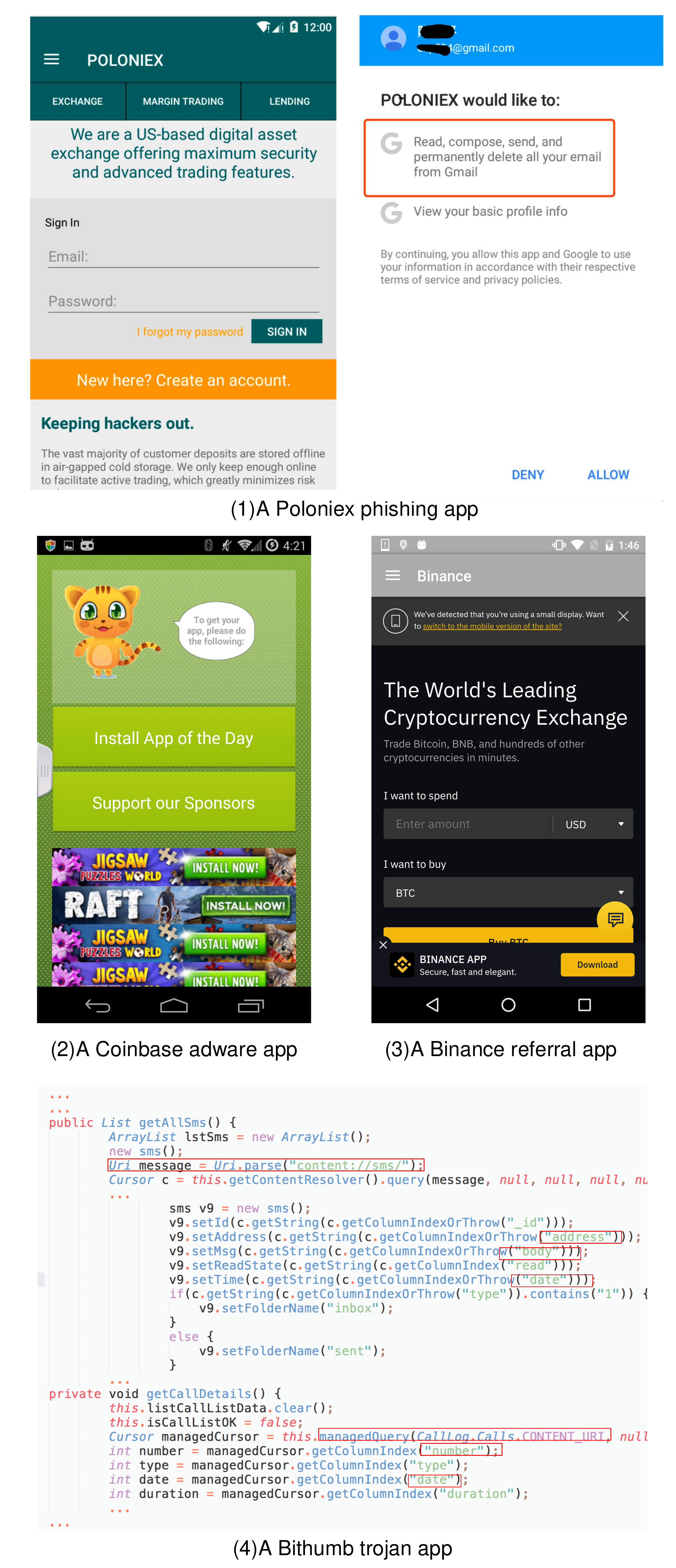}
\caption{Examples of Identified Fake Apps.}
\label{fig:appexample}
\end{figure}

\textbf{Examples.}
\textit{Figure~\ref{fig:appexample}} shows examples of the fake apps we identified. 
\textit{Figure~\ref{fig:appexample}(1)} shows a phishing app\footnote{SHA256:559f70db9f6e6741b59bdb2ad99f8ac53f5915e6ef3cab33522ce27cd9ccadb9} that targets \textit{Poloniex}. 
It fabricates a fake login screen and tricks users into typing in their Poloniex accounts. 
After that, it will continue to display a fake 2FA verification screen and ask for full email access to further steal users' email accounts.
Once success, attackers will get full access to users' Poloniex accounts and steal their money in a silent way.
\textit{Figure~\ref{fig:appexample}(2)} is a Coinbase adware\footnote{SHA256:91b523bdc7ffc7647b29d479b0e553f7d14f26d768960853eb8e9bb5d4493685} sample. It was repackaged from official Coinbase app, and embedded with aggressive ad libraries. 
During run-time, it requires users to install recommended apps, otherwise users cannot access to the main functionality of the app. However, most of the recommended apps are considered to be malicious. Furthermore, the app will push mobile ads during its running at background, which could lead to the unintentionally clicking of the advertisement. \textit{Figure~\ref{fig:appexample}(3)} is a referral app\footnote{SHA256:9efe538f464632b355086e06e6f07c785ccde94b5561647bd60628c3b2a261e7} that targets \textit{Binance}.
It simply implements a webview and connects to the referral link \textit{https://www.binance.com/?ref=20270961}. Attackers will benefit from users who register from this links. The benefit is usually a portion of commission fee, depending on the referral rules of different exchanges\footnote{https://www.binance.com/en/activity/referral}. \textit{Figure~\ref{fig:appexample}(4)} is a code snippet of a Bithumb trojan app\footnote{SHA256:e50e1dacbe6dcd8653c47c767d664860907e53adc55af5da056282bfabc898ca}. 
As highlighted in the decompiled code, it will collect users' text messages, contracts, and call logs secretly, and then upload them to \textit{http://bithumbinback.pro/}, which was the attacker's private server. Moreover, it monitors the device's incoming calls and messages at background.

\vspace{0.1in}
\noindent\fbox{
	\parbox{0.95\linewidth}{
		\textbf{Answer to RQ1.2:} 
Fake exchange apps are also prevalent in the ecosystem. Over 38 exchanges are targeted by 323 fake apps. A number of them show malicious behaviors and pose great security threats to mobile users.
}}

\section{Understanding the Attackers}
\label{sec:attackers}

\begin{figure*}[t]
\centering
\includegraphics[width=1.0\textwidth]{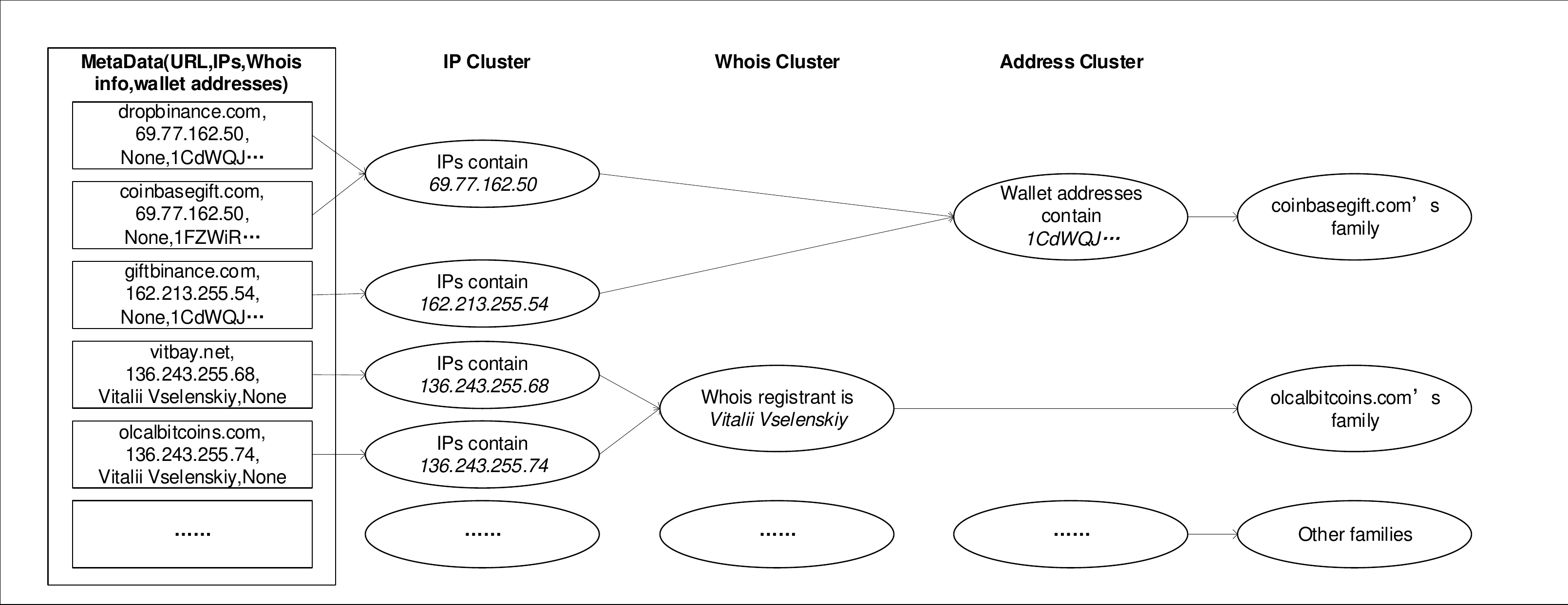}
\caption{A Three-phase Domain Clustering.}
\label{fig:domaincluster}
\end{figure*}

Our previous exploration suggests that exchange scams are prevalent in the ecosystem. In this section, we further investigate the relationship between these scams, in order to understand the attackers behind them. 
We first correlate the scam domains based on the information we collected, then we group fake apps based on code similarity and developer signatures. At last, we further identify the relationship between scam domains and fake apps.

\subsection{The Relation of the Scam Domains}

\subsubsection{Approach.}
We use a three-phase clustering approach to characterize their relationship, as shown in Figure~\ref{fig:domaincluster}.

\textbf{IP clustering.}
For each scam domain, we first resort to urlscan\footnote{https://urlscan.io/} to collect all the related IP addresses by searching its history resolve IPs (Passive DNS). 
We have collected $1,348$ unique IP addresses that related to the $1,534$ scam domains. Note that one domains may correspond to multiple IP addresses, and several domains may share the same IP addresses.
In Section~\ref{subsec:domainlabel}, we collected the domains that uses parking services and their IPs. We further remove IPs related to parking service in case they affect the cluster result. Then we group the domains based on IP addresses. During grouping, we also find other IPs related to parking services or domain hosting services due to their uncommon cluster sizes, and we remove them too. 
At last, we have remained $1,215$ IP addresses that we believe were used for malicious purposes.
After this step, we have 76 clusters in total, with 580 domains in the clusters. Note that, the remaining 1015 domains are isolated in this step.

\textbf{Whois Clustering.}
Whois information usually contains some personal data of the domain holders, which may help identify the domain groups held by each attacker\cite{quinkert2019s,khan2015every}. 
Only 266 domains (16.7\%) in our datasets have corresponding valid Whois information\footnote{The Whois information of some domains is replaced with the words like 'GDPR Masked' due to privacy consideration, thus we only consider Whois information that contains a valid distinguishable registrant name.}. 
Among 266 domains, there are 51 unique Whois information and 11 of them are shared by 226 domains. 
This results in 6 new clusters, and 2 clusters in step 1 are combined.
Therefore, after this step, we further cluster the domains to 81 clusters (including 644 domains).

\textbf{Blockchain Address Clustering.}
\label{sec:blockchainaddr}
As some identified domains are used for trading scams, and they have embedded the scam blockchain addresses in the corresponding webpages. 
Thus we further analyzed the crawled HTML webpage, and use regular expressions and checksum to identify blockchain addresses. Table~\ref{tab:btcregex} shows examples of regular expressions we used to identify Bitcoin and Ethereum addresses, respectively. 
We have identified 182 blockchain addresses in total, acrossing 6 kinds of Cryptocurrencies, including Ethereum, Bitcoin, XRP, Tron, NEO and Binance Coin, as shown in Table~\ref{tab:addresssum}. More specifically, we have 66 Bitcoin scam addresses and 111 Ethereum scam addresses,
Then, we group the domains based on these addresses, and achive the final clustering results.

\begin{table}[]
\caption{Examples of regular expressions we use to identify blockchain addresses.}
\begin{tabular}{@{}cc@{}}
\toprule
Cryptocurrencies & Regular expression \\ \midrule
Bitcoin & {}(bc1|{[}13{]}){[}a-zA-HJ-NP-Z0-9{]}\{25,39\} \\
Ethereum & 0x{[}a-fA-F0-9{]}\{40\} \\ \bottomrule
\end{tabular}
\label{tab:btcregex}
\end{table}

\begin{table}[]
\caption{A summary of blockchain addresses we got from scam domains.}
\resizebox{\linewidth}{!}{

\begin{tabular}{@{}ccccc@{}}
\toprule
Blockchain & \begin{tabular}[c]{@{}c@{}}\# target\\ exchanges\end{tabular} & Target exchanges & \# domains & \# addr \\ \midrule
Ethereum & 19 & \begin{tabular}[c]{@{}c@{}}Binance,Bibox,OKEx,Cobinhood\\ ,Coinbase,BitMEX,...\end{tabular} & 138 & 111 \\
Bitcoin & 7 & \begin{tabular}[c]{@{}c@{}}Binance,Coinbase,Huobi,Kraken,\\ OKEx,BitMEX,Yobit\end{tabular} & 85 & 66 \\
XRP & 2 & Binance,Kucoin & 5 & 2 \\
Tron & 2 & Poloniex,Kucoin & 5 & 1 \\
NEO & 2 & Poloniex,Kucoin & 3 & 1 \\
Binance Coin & 1 & Binance & 1 & 1 \\ \bottomrule
\end{tabular}
}
\label{tab:addresssum}
\end{table}

\subsubsection{Results.}
\label{subsec:clusterresult}
\begin{figure}[t]
\includegraphics[width=0.45\textwidth]{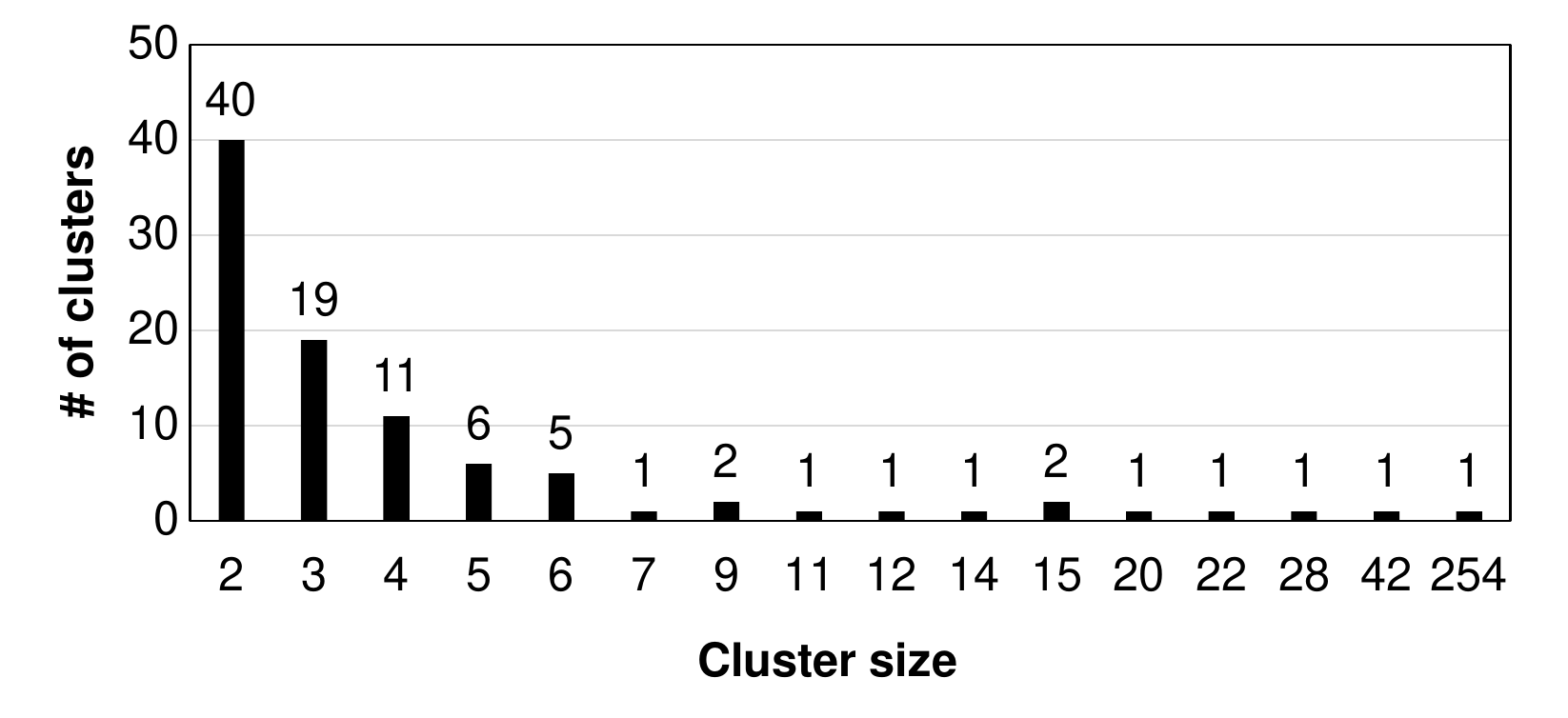}
\caption{The distribution of cluster size.}
\label{fig:clustersize}
\end{figure}

At last, we have obtained 94 clusters, with 699 domains (43.8\%) in total.
Note that there are 896 isolated domains. 
The distribution of cluster size is shown in Figure~\ref{fig:clustersize}.
We can observe that most of the clusters are small clusters with size 2 or 3, and there are only 18 clusters with a size larger than 5.
Table~\ref{tab:topcluster} lists the top-15 clusters, which we have assigned each cluster a family name. 
This result suggests that: (1) \textit{some attackers have the tendency to created a large number of scam domains}. For example, the largest family in our dataset have created 254 scam domains, targeting 11 different exchanges; (2) \textit{attackers tend to use the same method to create the scam domains}, i.e., the scam category remains the same for most clusters. 
The reason might be that it is easier for them to reuse one method in creating multiple scam domains.

\begin{table*}[h]
\caption{Top-15 clusters ranked by the number of scam domains. Note that we have assigned a family name for each cluster by randomly picking one scam domain name.}
\resizebox{\linewidth}{!}{

\begin{tabular}{|c|c|c|c|c|c|c|c|}
\hline
Family & \# urls & \# exchanges & Target exchanges & Category & Shared IPs & Shared Whois & Shared addresses \\ \hline
olcalbitcoins.com & 254 & 11 & \begin{tabular}[c]{@{}c@{}}Binance,Livecoin,Wirex,\\ Bitbay,LocalBitcoins,...\end{tabular} & Referral Fraud & 136.243.255.0/24 & Vitalii Vselenskiy &  \\ \hline
coinchecl.com & 42 & 8 & \begin{tabular}[c]{@{}c@{}}Poloniex,Binance,CoinCheck,\\ Kraken,HitBTC,...\end{tabular} & Referral Fraud &  & xu shuaiwei &  \\ \hline
kralkem.com & 28 & 7 & \begin{tabular}[c]{@{}c@{}}Poloniex,Binance,Bittrex,\\ Kraken,KuCoin,...\end{tabular} & Phishing & 185.110.132.214 &  &  \\ \hline
coinbasegift.com & 22 & 3 & Coinbase,Binance,Kraken & Trading Scam & \begin{tabular}[c]{@{}c@{}}198.187.29.252,\\ 199.33.112.226,\\ 69.77.162.51,...\end{tabular} &  & \begin{tabular}[c]{@{}c@{}}1FZWiRH5zSwsaFY5gUFXVGML6NHsADngRp,\\ 19R9MWW88rZwivGWvvz15Ey9G7mpgJYesB,\\ 1CdWQJMiQF1uYbwKc7fb5VBb9JBrhykcNf\end{tabular} \\ \hline
bitma.io & 20 & 2 & BitMEX,BitMax & Referral Fraud & \begin{tabular}[c]{@{}c@{}}46.166.184.106,\\ 185.206.180.119\end{tabular} & Sun Wukong &  \\ \hline
virexapp.com & 15 & 5 & \begin{tabular}[c]{@{}c@{}}Wirex,BitMEX,Bitfinex,\\ Kraken,HitBTC\end{tabular} & Referral Fraud,Scam & 77.78.104.3 &  &  \\ \hline
bitbai.net & 15 & 3 & Yobit,Bitbay,Bitforex & Referral Fraud & \begin{tabular}[c]{@{}c@{}}212.91.7.33,\\ 185.253.212.22\end{tabular} &  &  \\ \hline
deribiy.com & 14 & 2 & Huobi,Deribit & Referral Fraud & 185.182.56.12 &  &  \\ \hline
bitpannda.com & 12 & 2 & Bitpanda,Coinmama & Referral Fraud & 78.109.174.110 &  &  \\ \hline
yobitr.net & 11 & 5 & \begin{tabular}[c]{@{}c@{}}Yobit,Binance,Kraken,\\ Hitbtc,Bittrex\end{tabular} & Phishing & 185.110.132.216 & Sergei Nesmiyanov &  \\ \hline
kueoin.com & 9 & 5 & \begin{tabular}[c]{@{}c@{}}Poloniex,BitMEX,Bitfinex,\\ KuCoin,Bittrex\end{tabular} & Phishing & 5.45.65.239 &  & \\ \hline
paxfulverify.com & 9 & 3 & Yobit,LocalBitcoins,Paxful & Phishing & \begin{tabular}[c]{@{}c@{}}204.93.160.0/19,\\ 198.38.82.0/24,...\end{tabular} &  &  \\ \hline
binance-presents.fund & 7 & 1 & Binance & Trading Scam & 162.144.100.203  &  &  \begin{tabular}[c]{@{}c@{}}1Mn386ue8o3mW9866octLNP8HFqcYsphJC,\\ 0x11775A106157a283873A81E8Ec58394b8d568E06\end{tabular}\\ \hline
loginviet-binance.com & 6 & 1 & Binance & Phishing & \begin{tabular}[c]{@{}c@{}}198.187.29.106,\\ 198.54.116.199,...\end{tabular} & Taraku Apostrof &  \\ \hline
giveaway-coinbase.top & 6 & 1 & Coinbase & Trading Scam & 181.215.237.183 &  & \begin{tabular}[c]{@{}c@{}}0xDF50C2DA0a52f2a3a231eD38fA1B79Ad97ab9563,\\ 0xfb5e36B888bc15528b6Bd42fe0B1b2aF62693eB9\end{tabular} \\ \hline
\end{tabular}
}
\label{tab:topcluster}
\end{table*}

\subsection{The Relation of the Fake Apps}

\subsubsection{Approach.}
We group fake apps based on both developer signatures and code similarity.

\textbf{Clustering based on developer signatures.}
Android system uses developer certificates to identify the authorship of apps.
In our dataset, we have extracted 206 unique developer signatures in total.
Note that, some fake apps may use the common signatures that widely known in our community. For example, Android framework has provided four common keys. Thus, we further analyze these signatures and remove a common Android framework signature '61ed377e85d386a8dfee\-6b864bd85b0bfaa5af81' (related to 6 exchanges and 9 fake apps).
We thus have 205 unique developer signatures in total.

\textbf{Clustering based on code similarity.}
Previous work suggested that malicious developers always reuse the code to generated apps. Thus, we further measure the code-level similarity of these fake apps. Here, we take advantage of SimiDroid~\cite{li2017simidroid}, a tool that provides comprehensive pairwise comparison to understand the similarity among apps. 
We perform pair-wise comparison to calculate method-level similarity of all the apps we collected.Apps with similarity higher than 80\%\footnote{The threshold is selected empirically based on previous studies~\cite{li2019rebooting}.} will be classified into a same cluster. 

\subsubsection{Results.}
For the signature-level clustering, we observe that 60 signatures were reused by fake apps, with 169 apps in total. The other remaining 145 signatures only have one corresponding app each\footnote{There are also 9 fake apps signed by the Android framework common signature.}.
Table~\ref{tab:sigcluster} shows the result.
Fake apps in signature-level clusters account for over 52\% of all the fake apps we identified.
As to the code-level clustering, we have clustered 34 groups, including 270 fake apps (83.6\%), with only 53 isolated apps.
Table~\ref{tab:codecluster} shows the top-5 code similarity clustering groups.

Results show that both developer signatures and code similarity can help identify attacker groups. 
We further study the relations of the signature-same clusters and code-similar clusters. 
Based on the code similarity clusters, we combine the clusters that contain same signatures and then add signature clusters that are not in the similarity clusters. 
At last, we combine 9 similarity clusters into 3 clusters and add two signature clusters. 
We finally have grouped 30 app clusters, with 275 apps (85.1\%) in total (with only 48 isolated apps). 

We further sampled apps from each cluster for manually inspection, and we have the following observations. 
First, fake apps signed by the same certificate are usually with high code similarity, indicating that they share the similar malicious behaviors and purposes. Second, similar to the scam domains, to reduce development cost, quite a few attackers prefer to use easy-to-use visual programming platforms like App Inventor\footnote{https://appinventor.mit.edu/} or AppsGeyser(https://appsgeyser.com/) to develop their forged apps with a specific template.

\begin{table}[t]
\caption{Top-10 fake app clusters (signature-level).}
\resizebox{\linewidth}{!}{

\begin{tabular}{@{}ccc@{}}
\toprule
Develeper signature(SHA1) &
  Target exchanges &
  \begin{tabular}[c]{@{}c@{}}\# apps\end{tabular} \\
  \midrule
2CB7E9064D1EC3852191B45F3645A02EF55105B9  & \begin{tabular}[c]{@{}c@{}}Kraken,Bitstamp,\\ Cryptopia,HitBTC\end{tabular} & 4 \\ \hline
21064B6D32EB94D49143FE23F06BD222C116B348  & \begin{tabular}[c]{@{}c@{}}Paxful,Coinmama,\\ Bitfinex\end{tabular}         & 3 \\ \hline
7CD76D9FEA4736AEFF636AD02512FFE625702FC6  & \begin{tabular}[c]{@{}c@{}}Cryptopia,Poloniex,\\ Bittrex\end{tabular}       & 3 \\ \hline
86DA54FDAD362FC78354C987E4337F762D37B702 & \begin{tabular}[c]{@{}c@{}}Bitfinex,Coinmama,\\ Bitstamp\end{tabular}       & 3 \\ \hline
E99A56A0F329F243CC2759317F07E94CDF9ACFA8 & \begin{tabular}[c]{@{}c@{}}Cryptopia,Poloniex,\\ Bittrex\end{tabular}       & 3 \\ \hline
7B927F47E2F99722846F9706E3B1CAD129E17D90 & Cryptopia,Bittrex                                                           & 5 \\ \hline
AF49696504D84B6BD15E3B505EC79049F45DCC73  & Bitfinex,Bitstamp                                                           & 3 \\ \hline
8DDD7A5D446A3FEAE270DA5BBC6A14186CD4843E  & CoinExchange,LocalBitcoins                                                  & 2 \\ \hline
7238E7D72F225EBCD660B0932E47B3197BCE1EB7  & LocalBitcoins,Poloniex                                                      & 2 \\  \hline
F16B1CD5DA076CEEEE8BB1523B25B63EC6FAA171  & CoinExchange                                                      & 9\\ \bottomrule
\end{tabular}
}
\label{tab:sigcluster}
\end{table}

\begin{table}[t]
\caption{Top-5 code similarity clusters which have most apps.}
\resizebox{\linewidth}{!}{
\begin{tabular}{@{}ccccc@{}}
\toprule
an app's SHA256 in fake app families &
  \begin{tabular}[c]{@{}c@{}}\# of apps in \\ same family\\ (\# of apps \\ reported by \\ VirusTotal)\end{tabular} &
  \begin{tabular}[c]{@{}c@{}}\# of \\ Target \\ exchanges\end{tabular} &
  Target exchanges &
  Reported malware types \\ \midrule
\begin{tabular}[c]{@{}c@{}}96348ed94d796d7c0f3459560ca499d-\\ adfb852c678ecc5ba3d3dfac0dcf261d3\end{tabular} &
  86(71) &
  18 &
  \begin{tabular}[c]{@{}c@{}}Bibox,Binance,BitMEX,\\ bitstamp,Bittrex,...\end{tabular} &
  Gray,Trojan \\
\begin{tabular}[c]{@{}c@{}}6cb9ab55b6d9dcc85c585546408de196-\\ 2391a49a66ec2afae39b208d29dc9d4a\end{tabular} &
  33(11) &
  15 &
  \begin{tabular}[c]{@{}c@{}}b2bx,Binance,bitfinex,\\ BitMEX,bitso,...\end{tabular} &
  Adware,Trojan \\
\begin{tabular}[c]{@{}c@{}}818c1a91dd50494adb01748da9f7c8b2-\\ 2a5464fb971e3c6085e977516705057d\end{tabular} &
  16(12) &
  9 &
  \begin{tabular}[c]{@{}c@{}}Binance,bitfinex,BitMEX,\\ bitstamp,Bittrex,...\end{tabular} &
  \begin{tabular}[c]{@{}c@{}}Repmalware,Riskware,\\ Trojan\end{tabular} \\
\begin{tabular}[c]{@{}c@{}}7f5dab9450ba1d7ae6da0747b24ab4c9-\\ 88cb09edecd34277c3b391349f6ca1a1\end{tabular} &
  15(3) &
  7 &
  \begin{tabular}[c]{@{}c@{}}Binance,bitforex,cryptopia,\\ etoro,hitbtc,...\end{tabular} &
  \begin{tabular}[c]{@{}c@{}}Adware,Gray,\\ Repmalware\end{tabular} \\
\begin{tabular}[c]{@{}c@{}}82d8edcf60fd265114280ba2824f34d2-\\ 9d86066be006275b4d8fa1a69c4db803\end{tabular} &
  10(0) &
  6 &
  \begin{tabular}[c]{@{}c@{}}Bittrex,cryptopia,hitbtc,\\ Kraken,Poloniex,yobit\end{tabular} &
   \\ \bottomrule
\end{tabular}
}
\label{tab:codecluster}
\end{table}

\subsection{The Tie between Scam Domains and Apps}
To investigate the tie between scam domains and fake apps, we further analyzed the connected URLs and domains of fake apps, to see whether they are overlapped with the scam domains we identified.
Therefore, we extract urls from the fake apps using Apkatshu\footnote{https://github.com/0xPwny/Apkatshu}, a popular tool for extracting urls, emails, and IP addresses from apk files.
We seek to investigate whether we could find some clues to link the scam domains and apps.
However, from the data we collected, we can only find 1 phishing url (xn--polonix-y8a.com) targeting at Poloniex.
This result suggested that, there is no clear evidence to link scam domains and the fake apps. 
The reason might be that, fake apps usually use the app name and user interface to infect unsuspicious users. 
As long as users have installed the fake apps, the malicious behaviors can be performed in either foreground (e.g., using the fake UIs) or background (e.g., stealthy behaviors), without the need of further using squatting domains to cheat users.

\vspace{0.1in}
\noindent\fbox{
	\parbox{0.95\linewidth}{
		\textbf{Answer to RQ2:} 
Our experiment results suggested that a number of the scams are controlled in groups, i.e., 43.8\% of the scam domains and 85.1\% of fake apps could be clustered into groups. This observation could help us identify and track the new scams in the future. For example, new domains that related to existing scam IP addresses and blockchain addresses are high suspicious to be malicious. The apps released by the known scam signatures should also be paid special attention to.}}

\section{Characterizing the Impacts}
\label{sec:impact}

In this section, we measure the impacts of cryptocurrency exchange scams from two ways. First, we trace the money flow of scam addresses, in order to quantifying the scale of the scams, i.e., the number of victims and the amount of financial loss. 
Second, we measure the presence of fake apps on major app markets, to see how many of them have penetrated to popular app markets to infect unsuspicious users.

\subsection{Money flow of scam addresses}

As aforementioned in Section~\ref{sec:blockchainaddr}, we extract 182 unique wallet addresses contained in the trading scam websites. Thus, we further analyze the transaction information related to these addresses to estimate the impact the scams. Note that, the financial loss we estimated here is a lower-bound of the whole ecosystem, considering that there are many scam domains that we are not able to directly investigate their impacts here.

\subsubsection{Overall Result}

We further retrieved all the transaction data related to these addresses. There are $1,713$ income transactions taken place and these scam address received a total number of 28.84 BTC, 1625.29 ETH, and other tokens, which is equivalent to roughly over 520K US dollars\footnote{Note that, this amount of money is calculated based on BTC and ETH's price at 2020-01-21(BTC:8625.16\$, ETH:167.25\$) and the same below}.

\begin{figure}[t]
\centering
\includegraphics[width=0.45\textwidth]{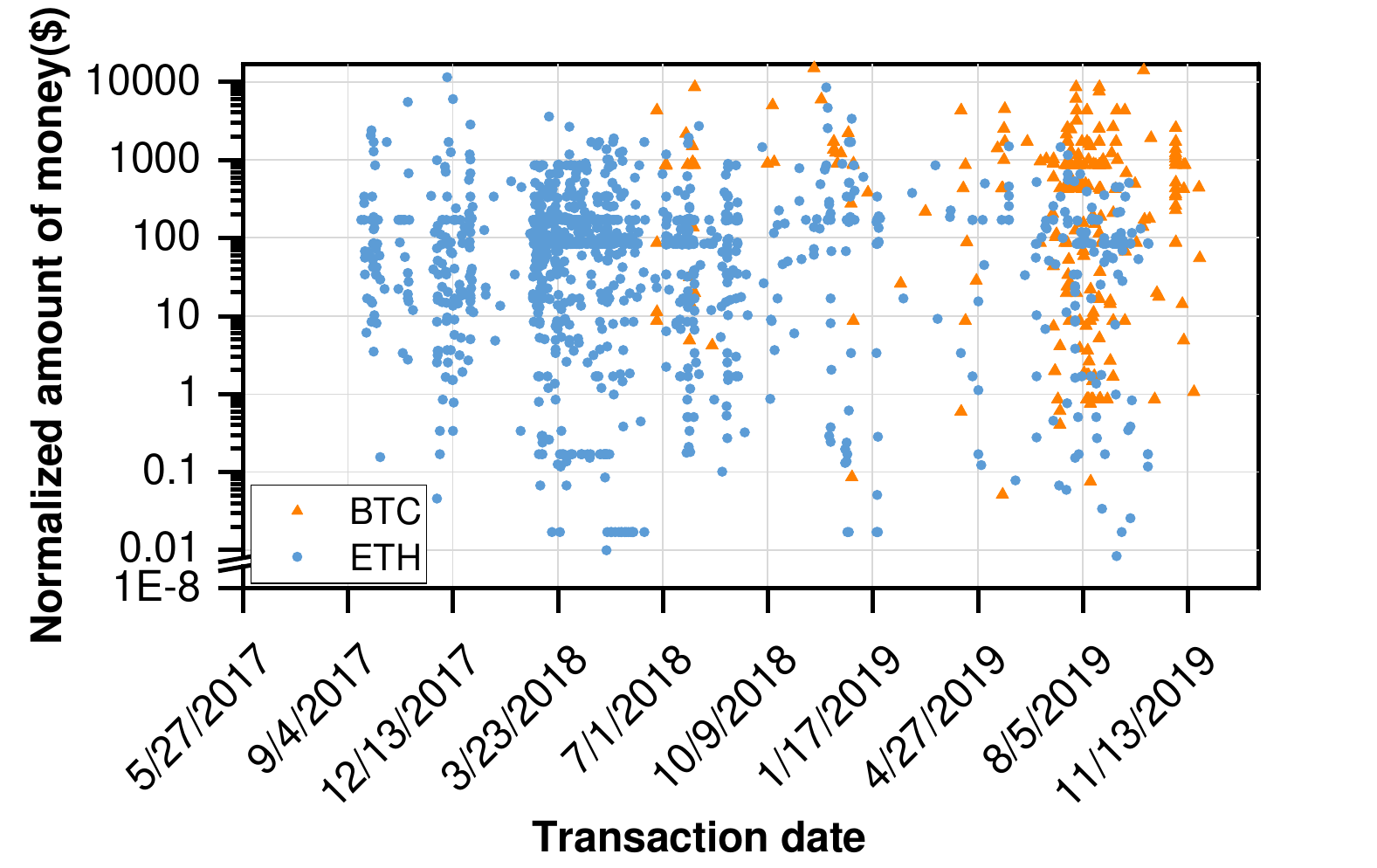}
\caption{The scatter diagram of two major cryptocurrencies' 1,659 transactions.}
\label{fig:addressdistribute}
\end{figure}

\textbf{Distribution.}
We analyze the distribution of BTC and ETH addresses' incoming transactions, as shown in Figure~\ref{fig:addressdistribute}. 
For the amount of money loss, over 41.8\% of the transactions are over 100 US Dollars. The largest transaction record took place on 2018-11-22 and the scam address \textit{1MpLjpT44A5y\-yRbtGG61rtpgwxdJB3onsB} received about 15K dollars in total. 

As to the transaction time, the first victim was deceived 0.99 Ethereum (167.18\$) on Sept 16th 2017 on the \textit{cobinhood.io} whose target is Cobinhood shortly after the exchange's ICO launch. It is interesting to observe that, ETH is popular before 2019, while BTC turns to be more prevalent after July of 2019.

\begin{figure}[htbp]
\centering
\includegraphics[width=0.45\textwidth]{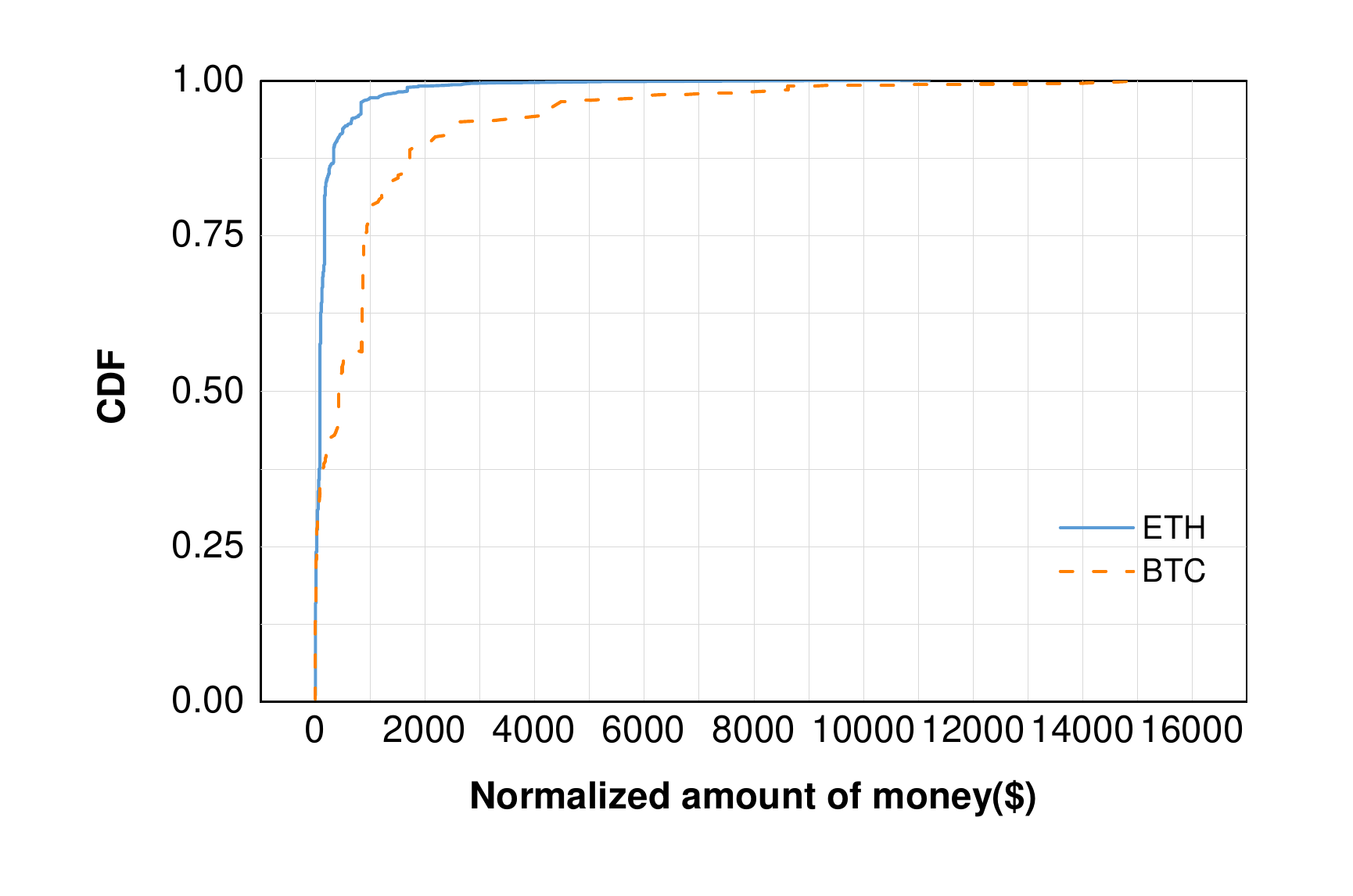}
\caption{The distribution of incoming transactions for each scam address.}
\label{fig:moneydistribute}
\end{figure}

\textbf{The most profitable addresses.}
On average, each scam address has received 9 transactions. 
While some addresses are more active than we expected, e.g., \textit{0x40949225c4a1745a9946f6aaf763241c082c\-b9ac} has received over 474 transactions. 
We further analyzed the amount of incoming transactions for BTC and ETH addresses, and the distribution is shown in Figure~\ref{fig:moneydistribute}. 
On average, each address has received 2941.05 Dollars.
Roughly 75\% of the BTC addresses have received less than 918.58 US Dollar equivalent tokens, and 80.04\% of ETH addresses have received less than 167.42 US Dollar equivalent tokens.
Table~\ref{tab:topaddress} lists the top-5 profitable addresses.
The largest one has received roughly 500 ETH, which is roughly equivalent to 83K US Dollars.

\begin{table*}[t]
\caption{Top-5 profitable addresses.}
\resizebox{\linewidth}{!}{
\begin{tabular}{@{}cccccc@{}}
\toprule
\begin{tabular}[c]{@{}c@{}}Target \\ exchange\end{tabular} &
  Scam domain &
  Scam address &
  
  \begin{tabular}[c]{@{}c@{}}\# total\\ incoming\\ transactions\end{tabular} &
  Total received &
  \begin{tabular}[c]{@{}c@{}}Current \\ value(\$)\end{tabular} \\ \midrule
Binance    & binancefree2018.droppages.com         & 0x40949225c4a1745a9946F6AAf763241c082cb9ac &   474  & 497.39 ETH & 83192.66 \\
shapeshift & shapishift.io,xn--hapeshit-ez9c2y.com & 0x3853ba76ec6ae97818e2d0e0839c9eda6c396690 &   140  & 309.13 ETH & 51702.10 \\
Coinbase   & coinbase-airdrop.top,coinbase-btc.xyz & 1MpLjpT44A5yyRbtGG61rtpgwxdJB3onsB         &   28  & 4.93 BTC   & 42537.83 \\
Binance    & dropbinance.com,giftbinance.com       & 1CdWQJMiQF1uYbwKc7fb5VBb9JBrhykcNf         &   13  & 4.43 BTC  & 38232.70 \\
Binance    & binance.claims                        & 13XzbaQV6k21yfbS5WDkzwSPkAxQ1AsbQ3         &   14  & 1.96 BTC  & 16950.98 \\ \bottomrule
\end{tabular}
}
\label{tab:topaddress}
\end{table*}

\begin{table*}[t]
\caption{Top-5 profitable families.}
\resizebox{\linewidth}{!}{
\begin{tabular}{@{}cccccccc@{}}
\toprule
\begin{tabular}[c]{@{}c@{}}Target \\ exchanges\end{tabular} &
  Family &
  \# domains &
  \# addresses &
  Addresses &
  \begin{tabular}[c]{@{}c@{}}\# total \\ incoming \\  transactions\end{tabular}
    &
  Total received &
  \begin{tabular}[c]{@{}c@{}}Current \\ value(\$)\end{tabular} \\ \midrule
\begin{tabular}[c]{@{}c@{}}Binance,\\ Coinbase,\\ Kraken\end{tabular} &
  coinbasegift.com &
  22 &
  18 &
  \begin{tabular}[c]{@{}c@{}}1FZWiRH5zSwsaFY5gUFXVGML6NHsADngRp,\\ 19R9MWW88rZwivGWvvz15Ey9G7mpgJYesB,\\ 1CdWQJMiQF1uYbwKc7fb5VBb9JBrhykcNf,...\end{tabular} &
  65 &
  8.25 BTC &
  71128.11 \\ \midrule
Coinbase &
  coinbase-btc.xyz &
  2 &
  1 &
  1MpLjpT44A5yyRbtGG61rtpgwxdJB3onsB &
  28 &
  4.93 BTC &
  42537.83 \\ \midrule
Binance &
  binance.updog.co &
  4 &
  2 &
  \begin{tabular}[c]{@{}c@{}}0x76bb5b6177096b337c79F2f948Aa08b0db5f5211,\\ 13XzbaQV6k21yfbS5WDkzwSPkAxQ1AsbQ3\end{tabular} &
  56 &
  \begin{tabular}[c]{@{}c@{}}35.79 ETH,\\ 1.97 BTC\end{tabular} &
  23028.14 \\ \midrule
\begin{tabular}[c]{@{}c@{}}Bithumb,\\ Huobi\end{tabular} &
  huobiglobal.ltd &
  3 &
  1 &
  0xe2e4B53A1324F5a7368724eA73e532c626517f19 &
  107 &
  60.48 ETH &
  10411.29 \\ \midrule
Binance &
  binance-presents.fund &
  7 &
  4 &
  \begin{tabular}[c]{@{}c@{}}0x11775A106157a283873A81E8Ec58394b8d568E06,\\ 1Mn386ue8o3mW9866octLNP8HFqcYsphJC,...\end{tabular} &
  29 &
  \begin{tabular}[c]{@{}c@{}}20.70 ETH,\\ 0.62 BTC\end{tabular} &
  8794.36 \\ \bottomrule
\end{tabular}
}
\label{tab:topfamily}
\end{table*}

\textbf{Scam Families.}
We further analyze the scam families we identified in Section~\ref{subsec:clusterresult}. 
Among the 36 families that have at least one address, 35 of them are trading scams and the remaining one is a phishing family. They have 68 addresses in total, while the other 114 addresses are isolated. Note that, 25 families (69.4\% of the families that have blockchain addresses) have only one corresponding blockchain address. The top 5 profitable families are listed in Table~\ref{tab:topfamily}. The family \textit{'coinbasegift.com'} is most profitable, and it has received over $70,000$ equivalent US dollars in BTC.

\begin{figure*}[htbp]\centering 

\subfigure[BTC addresses' fund transfer flow.]{ 
\begin{minipage}{8cm}\centering
\includegraphics[width = 1\textwidth]{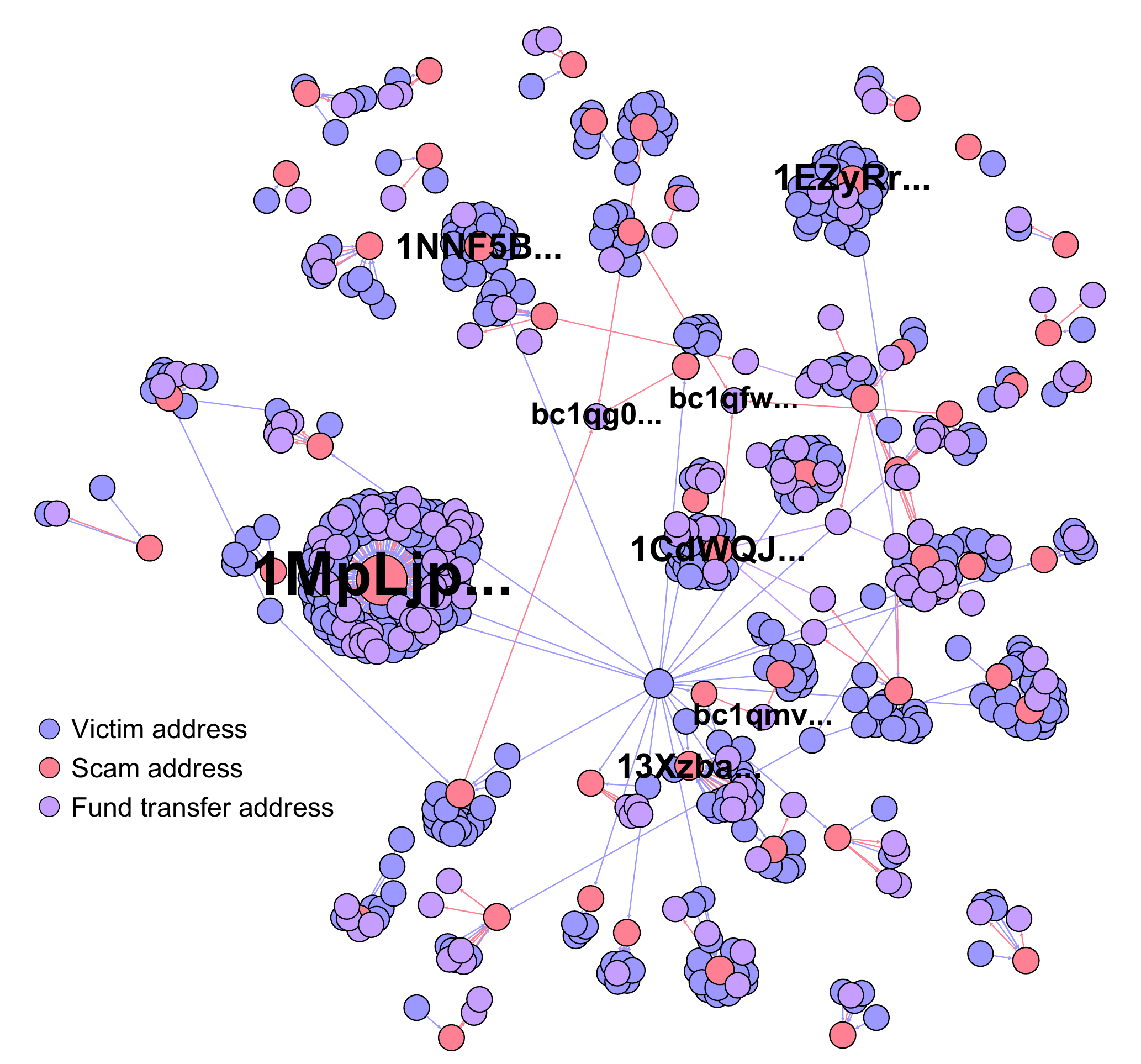} 
\end{minipage}}
\subfigure[ETH addresses' fund transfer flow.]{ 
\begin{minipage}{8cm}
\centering 
\includegraphics[width = 1\textwidth]{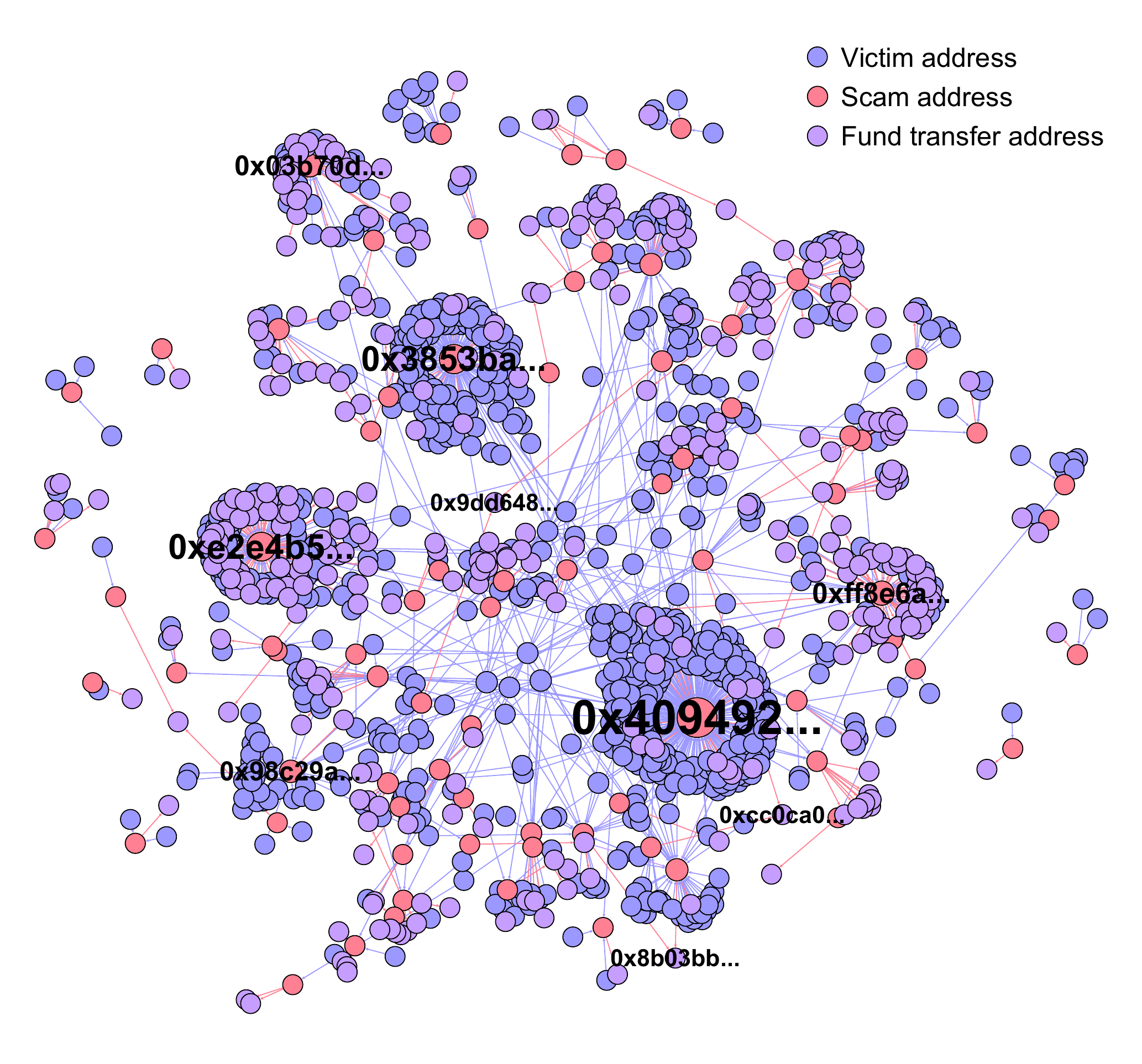} 
\end{minipage}
}
\caption{Fund transfer diagram of two major cryptocurrencies.}
\label{fig:fundtransferflow}
\end{figure*}

\subsubsection{Money Flow}

We further attempt to identify the relations between scam addresses by sorting out the money flows. 
We first label the addresses in the money flow into three categories:
1) \textit{the scam addresses}, the addresses that we extracted from the scam websites. Note that not all the addresses we found have transactions records, thus we remove the silent scam addresses during the money flow analysis.
2) \textit{the victim addresses}, which have ever transferred money to the scam addresses and did not receive money from scam addresses; 
3) \textit{the fund transfer addresses}, which were used to transfer money originated from the scam addresses.
Note that BTC's change addresses\cite{Change} are also a part of money flow, we consider them as fund transfer addresses. Figure~\ref{fig:fundtransferflow} shows the money flow of two major cryptocurrencies, BTC and ETH, respectively.

\textbf{Scam analysis.}
There are 1,320 victim addresses (470 in BTC and 850 in ETH) and 132 scam addresses (53 in BTC and 79 in ETH). 
In victim addresses, we find many of them have transferred money to multiple scam addresses. For example, the address \textit{0xfbb1b73c4f0bda\-4f67dca266ce6ef42f520fbb98} transferred money to 10 scam addresses with 68 transactions. 

On average, each BTC scam address is related to 9 victim addresses while each ETH scam address links to about 11, which may suggests that ETH-based scams have a slightly higher success rate.

\textbf{Fund transfer analysis.}
There are 518 fund transfer addresses (165 in BTC and 353 in ETH), which are far more than the scam addresses. We further studied their relations. We found 28 addresses share 13 fund transfer addresses. Table~\ref{tab:toptransfer} shows the top-5 of them. It is interesting to see that scam family \textit{coinbasegift.com} accounts for most of the top shared fund transfer addresses. Considering that this family also has mutiple scam addresses and transactions among scam addresses\footnote{\textit{12u54UVjvwVmzxNBBjTHtC6dgsVeZdr6RR} to \textit{bc1qerc6yxzre8xrcdfjx4zkgprafde30lr89vpd5s} and \textit{bc1qerc6yxzre8xrcdfjx4zkgprafde30lr89vpd5s} to \textit{bc1qmjwhdlz2wvdfrpmrgeydkej5eyv9djjqvsp3lz}}, this team of attackers is likely to carry out a careful plan to avoid tracking and we found only the tip of the iceberg. Besides, we found most of the fund transfer addresses have transferred all the tokens they received, suggesting that most of the attackers have transfer money through a chain of addresses. To better cover their tracks, attackers may further use the mixing service~\cite{Mixing} to achieve the purpose of money laundering.

\begin{table*}[t]
\caption{The top-5 fund transfer addresses ordered by the number of shared scam addresses.}
\resizebox{\linewidth}{!}{
\begin{tabular}{@{}cccccc@{}}
\toprule
Scam addresses &
  Scam family &
  Fund transfer address &
  \begin{tabular}[c]{@{}c@{}}\# related \\ scam \\ addresses\end{tabular} &
  \begin{tabular}[c]{@{}c@{}}\# of related \\ transactions\end{tabular} &
  \begin{tabular}[c]{@{}c@{}}Total \\ received(\$)\end{tabular} \\ \midrule
\begin{tabular}[c]{@{}c@{}}13tsX2zBiPz3P2nt5HgyFKxXTQFWRqXEuj,\\ 1FGZE75bUCHkoEcaoQLRzBuPYB9NA8XRCQ,\\ 1FZWiRH5zSwsaFY5gUFXVGML6NHsADngRp\end{tabular} &
  \begin{tabular}[c]{@{}c@{}}2 isolated ,\\ 1 in coinbasegift.com\end{tabular} &
  bc1qg09hzxsprzhh3fqdhcf6qtg9kcvcvwrp6nuyly &
  3 &
  3 &
  10162.78 \\ \midrule
\begin{tabular}[c]{@{}c@{}}1CdWQJMiQF1uYbwKc7fb5VBb9JBrhykcNf,\\ 1Lkakee2QGSQ92uNBUCUD1LaL4RKQTobLG,\\ 1BdencTWBaDrxpVBK7PPDtb9cot5Ns8D1T\end{tabular} &
  coinbasegift.com &
  bc1qfw3660gw5xv0t9p594hlq2xkmlkz0gmzley003 &
  3 &
  3 &
  5458.23 \\ \midrule
\begin{tabular}[c]{@{}c@{}}0x1363077895b20ae90f80794ce4e575559517d033,\\ 0x915c95415d3449212fd0991ccf5eb42864118ec9\end{tabular} &
  2 isolated &
  0x8b03bbe38069a34d1ab6db2f545f6cb8cd2d6a1e &
  2 &
  4 &
  11481.87 \\ \midrule
\begin{tabular}[c]{@{}c@{}}0x2784574e2405a7d3be1259b5f00412ae652018f4,\\ 0x3e9163816b073c2ce425c99e68ba8ae7caaec067\end{tabular} &
  \begin{tabular}[c]{@{}c@{}}1 isolated, \\ 1 in  binanceth.net\end{tabular} &
  0x9dd648a58cb8d2b5fbf937b863c627ba747dbf12 &
  2 &
  2 &
  5763.00 \\ \midrule
\begin{tabular}[c]{@{}c@{}}1NuZ4rxsQPU4izgtkScs793Uxx2c6ADRQo,\\ 16wd9B1LiXmTNf9hxQyb3Q9fbVHzP3NvSV\end{tabular} &
  \begin{tabular}[c]{@{}c@{}}1 in coinbasegift.com, \\ 1 in win-binance.com\end{tabular} &
  bc1qmvpmfglf9wk4wchucjp7gdhk7gv3wny4vm7z37 &
  2 &
  2 &
  3704.43 \\ \bottomrule
\end{tabular}
}
\label{tab:toptransfer}
\end{table*}

\begin{figure}[htbp]
\centering
\includegraphics[width=0.45\textwidth]{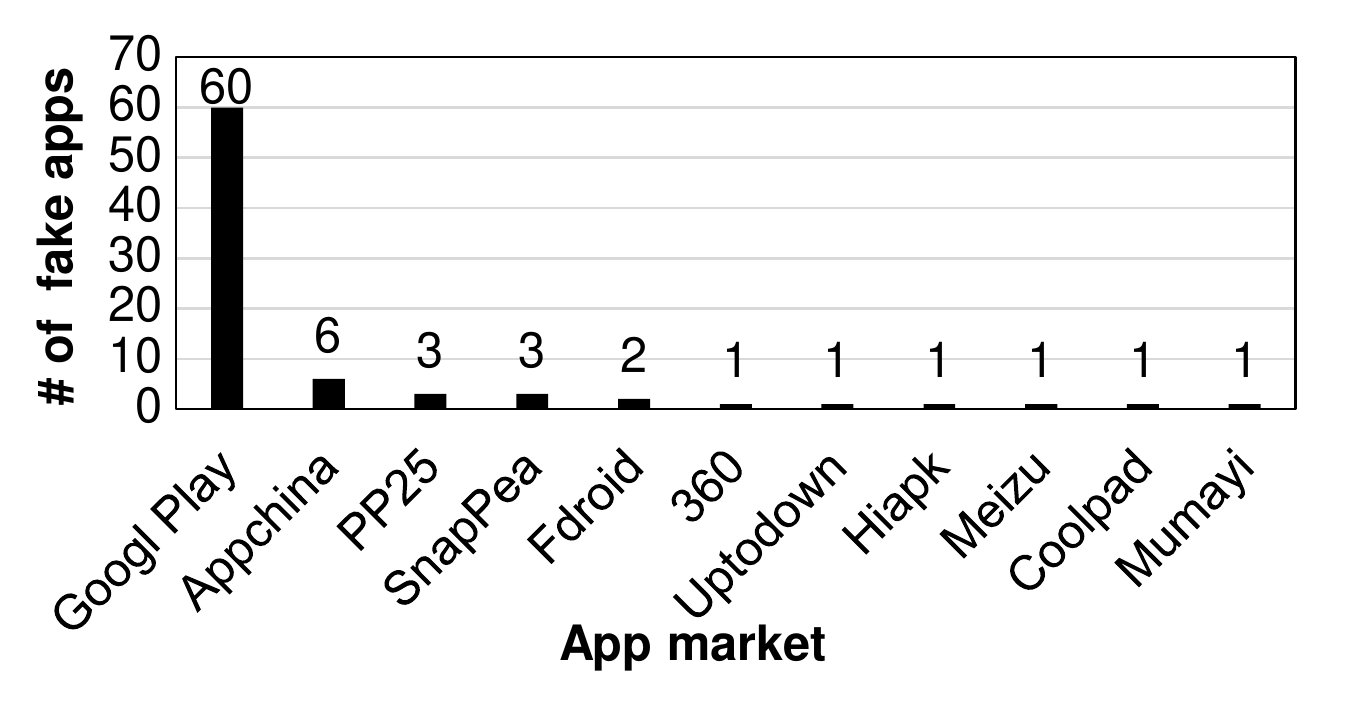}
\caption{The app market distribution of fake apps.}
\label{fig:appmarket}
\end{figure}

\subsection{Scams in Major App Markets}
As we have identified over 300 fake apps, we further analyze whether these apps have penetrated into major app markets. 

Although we crawled fake apps from Koodous, it does not contain app source information. Thus we resort to
Janus\footnote{https://appscan.io} and Androzoo\footnote{https://androzoo.uni.lu/}, two major app repositories to track fake apps' evidence in app markets. 
Among the 323 fake apps, over 66 (20.4\%) of them have been found in major app markets, as shown in Figure~\ref{fig:appmarket}.
Obviously, the official market -- Google Play, is the first target, with 60 fake apps in total.
Other third-party app markets, have hosted one or more fake apps. 
Note that, as we cannot get the download information of these fake apps\footnote{Most of them were removed from app markets}, we are not able to estimate the overall number of the victims here. 
Nevertheless, this result suggests that, existing app security check mechanisms deployed on app markets are not able to identify these fake apps effectively, which may affect many unsuspicious users.

\vspace{0.1in}
\noindent\fbox{
	\parbox{0.95\linewidth}{
		\textbf{Answer to RQ3:} 
Our experiment results suggested that there are about 1700 victims been deceived, with the amount scammed up to 520k dollars in our dataset. And although attackers' groups can be identified, they used multiple fund transfer addresses and mixing services to hide their tracks. On the other side, attackers have the ability to bypass the security check of the app markets and distribute their fake apps to markets, which exposes great threat to the community.
}}

\section{Implications and Limitations}

\subsection{Implications}
Our observations are of key importance to stakeholders in the blockchain community.
First, considering the large number of scam domains, fake apps, and blockchain addresses we discovered, the governance of the ecosystem needs to be improved. 
Second, considering most of the cryptocurrency exchanges are suffering from a growing number of scam attacks, our community should apply detection methods like we used in this paper to identify such scams and prevent users from being cheated by them. A growing and up-to-date scam database is also needed. 
Third, as we observed in this paper, many scams have strong relations and we could classify them into clusters.  This observation could guide us to identify new scams and raise alarms when new related domains found. 

\subsection{Limitations}
Our study carries some limitations. In several cases,
First, the methods and techniques we used in this paper are old-fashioned, i.e., typosquatting generation and fake app detection, and we also rely on some manually efforts in the study. Nevertheless, we have identify a number of scams and most of them have not been revealed to the community.
Some advanced techniques (e.g., machine learning techniques) could be used to identify and classify the scams.
Second, in this paper, we are only focused on the exchange scams. However, a number of scams may target cryptocurrency wallets and tokens. Thus, a promising future research direction is to study the scams in the other parts of the blockchain ecosystem.
Third, due to the limitation of dataset, we did not study the distribution channels of the scams, i.e., how do they get to users. The future direction might be investigating the advertisements of scams in social networking platforms (e.g., Facebook and Twitter).

\section{Concluding Remarks}

In this paper, we present the first systematic study of cryptocurrency exchange scams. We have created a dataset of over 1,500 scam domains and over 300 fake apps, and shared it to the community to boost future related research. We characterized the types and behaviors of scam domains and apps, and revealed that a majority of the scams were controlled by a small group of attackers. We further identified 183 blockchain addresses related to such attackers, and analyzed impacts of them.  
Our observations are of key importance to stakeholders in the blockchain community, and demonstrate the urgency to identify and prevent blockchain scams.

\section*{Acknowledgment}
This work was supported by the National Key Research and Development Program of China (No. 2018YFB0803603) and the National Natural Science Foundation of China (No. 61702045).

\balance
\bibliographystyle{abbrv}
\bibliography{sigproc} 

\begin{thebibliography}{10}

\bibitem{attacks2018}
{Group-IB: 14 cyber attacks on crypto exchanges resulted in a loss of \$882
  million}, 2018.
\newblock
  \url{http://securityaffairs.co/wordpress/77213/hacking/cyber-attacks-crypto-exchanges.html}.

\bibitem{binancebreach}
{Binance Security Breach Update}, 2019.
\newblock \url{https://www.binance.com/en/support/articles/360028031711}.

\bibitem{Change}
{Change}, 2019.
\newblock \url{https://en.bitcoin.it/wiki/Change}.

\bibitem{Mixing}
{Cryptocurrency mixing service}, 2019.
\newblock \url{https://en.wikipedia.org/wiki/Cryptocurrency_tumbler}.

\bibitem{dnstwist}
{Domain name permutation engine for detecting typo squatting, phishing and
  corporate espionage}, 2019.
\newblock \url{https://github.com/elceef/dnstwist}.

\bibitem{coinhousephishing}
{French Crypto Exchange Coinhouse Suffers Phishing Attack, User Names and
  Emails Accessed}, 2019.
\newblock
  \url{https://www.cryptoglobe.com/latest/2019/09/french-crypto-exchange-coinhouse-suffers-phishing-attack-user-names-and-\\emails-compromised/}.

\bibitem{listofcrypto}
{List of cryptocurrencies}, 2019.
\newblock \url{https://en.wikipedia.org/wiki/List\_of\_cryptocurrencies}.

\bibitem{URLCrazy}
{URLCrazy}, 2019.
\newblock \url{https://github.com/urbanadventurer/urlcrazy}.

\bibitem{agten2015seven}
P.~Agten, W.~Joosen, F.~Piessens, and N.~Nikiforakis.
\newblock Seven months' worth of mistakes: A longitudinal study of
  typosquatting abuse.
\newblock In {\em Proceedings of the 22nd Network and Distributed System
  Security Symposium (NDSS 2015)}. Internet Society, 2015.

\bibitem{alrwais2014understanding}
S.~Alrwais, K.~Yuan, E.~Alowaisheq, Z.~Li, and X.~Wang.
\newblock Understanding the dark side of domain parking.
\newblock In {\em 23rd $\{$USENIX$\}$ Security Symposium ($\{$USENIX$\}$
  Security 14)}, pages 207--222, 2014.

\bibitem{atzei2017survey}
N.~Atzei, M.~Bartoletti, and T.~Cimoli.
\newblock A survey of attacks on ethereum smart contracts (sok).
\newblock In {\em International conference on principles of security and
  trust}, pages 164--186. Springer, 2017.

\bibitem{bartoletti2020dissecting}
M.~Bartoletti, S.~Carta, T.~Cimoli, and R.~Saia.
\newblock Dissecting ponzi schemes on ethereum: identification, analysis, and
  impact.
\newblock {\em Future Generation Computer Systems}, 102:259--277, 2020.

\bibitem{bartoletti2018data}
M.~Bartoletti, B.~Pes, and S.~Serusi.
\newblock Data mining for detecting bitcoin ponzi schemes.
\newblock In {\em 2018 Crypto Valley Conference on Blockchain Technology
  (CVCBT)}, pages 75--84. IEEE, 2018.

\bibitem{bissias2016analysis}
G.~Bissias, B.~N. Levine, A.~P. Ozisik, and G.~Andresen.
\newblock An analysis of attacks on blockchain consensus.
\newblock {\em arXiv preprint arXiv:1610.07985}, 2016.

\bibitem{chen2018understanding}
T.~Chen, Y.~Zhu, Z.~Li, J.~Chen, X.~Li, X.~Luo, X.~Lin, and X.~Zhange.
\newblock Understanding ethereum via graph analysis.
\newblock In {\em IEEE INFOCOM 2018-IEEE Conference on Computer
  Communications}, pages 1484--1492. IEEE, 2018.

\bibitem{chen2018detecting}
W.~Chen, Z.~Zheng, J.~Cui, E.~Ngai, P.~Zheng, and Y.~Zhou.
\newblock Detecting ponzi schemes on ethereum: Towards healthier blockchain
  technology.
\newblock In {\em Proceedings of the 2018 World Wide Web Conference}, pages
  1409--1418, 2018.

\bibitem{dnstwist1}
T.~Dam, L.~D. Klausner, D.~Buhov, and S.~Schrittwieser.
\newblock Large-scale analysis of pop-up scam on typosquatting urls.
\newblock In {\em Proceedings of the 14th International Conference on
  Availability, Reliability and Security}, pages 1--9, 2019.

\bibitem{EOSIO}
Y.~Huang, H.~Wang, L.~Wu, G.~Tyson, X.~Luo, R.~Zhang, X.~Liu, G.~Huang, and
  X.~Jiang.
\newblock Characterizing eosio blockchain.
\newblock {\em arXiv preprint arXiv:2002.05369}, 2020.

\bibitem{hurier2017euphony}
M.~Hurier, G.~Suarez-Tangil, S.~K. Dash, T.~F. Bissyand{\'e}, Y.~Le~Traon,
  J.~Klein, and L.~Cavallaro.
\newblock Euphony: Harmonious unification of cacophonous anti-virus vendor
  labels for android malware.
\newblock In {\em 2017 IEEE/ACM 14th International Conference on Mining
  Software Repositories (MSR)}, pages 425--435. IEEE, 2017.

\bibitem{khan2015every}
M.~T. Khan, X.~Huo, Z.~Li, and C.~Kanich.
\newblock Every second counts: Quantifying the negative externalities of
  cybercrime via typosquatting.
\newblock In {\em 2015 IEEE Symposium on Security and Privacy}, pages 135--150.
  IEEE, 2015.

\bibitem{kywe2014detecting}
S.~M. Kywe, Y.~Li, R.~H. Deng, and J.~Hong.
\newblock Detecting camouflaged applications on mobile application markets.
\newblock In {\em International Conference on Information Security and
  Cryptology}, pages 241--254. Springer, 2014.

\bibitem{li2014zoom}
B.~Li, P.~Singh, and Q.~Wang.
\newblock Zoom in ios clones: Examining the antecedents and consequences of
  mobile app copycats.
\newblock 2014.

\bibitem{li2017simidroid}
L.~Li, T.~F. Bissyand{\'e}, and J.~Klein.
\newblock Simidroid: Identifying and explaining similarities in android apps.
\newblock In {\em 2017 IEEE Trustcom/BigDataSE/ICESS}, pages 136--143. IEEE,
  2017.

\bibitem{li2019rebooting}
L.~Li, T.~F. Bissyand{\'e}, and J.~Klein.
\newblock Rebooting research on detecting repackaged android apps: Literature
  review and benchmark.
\newblock {\em IEEE Transactions on Software Engineering}, 2019.

\bibitem{metcalf2014domain}
L.~Metcalf and J.~Spring.
\newblock Domain parking: Not as malicious as expected.
\newblock Technical report, CARNEGIE-MELLON UNIV PITTSBURGH PA SOFTWARE
  ENGINEERING INST, 2014.

\bibitem{quinkert2019s}
F.~Quinkert, T.~Lauinger, W.~Robertson, E.~Kirda, and T.~Holz.
\newblock It's not what it looks like: Measuring attacks and defensive
  registrations of homograph domains.
\newblock In {\em 2019 IEEE Conference on Communications and Network Security
  (CNS)}, pages 259--267. IEEE, 2019.

\bibitem{szurdi2014long}
J.~Szurdi, B.~Kocso, G.~Cseh, J.~Spring, M.~Felegyhazi, and C.~Kanich.
\newblock The long “taile” of typosquatting domain names.
\newblock In {\em 23rd $\{$USENIX$\}$ Security Symposium ($\{$USENIX$\}$
  Security 14)}, pages 191--206, 2014.

\bibitem{tang2019large}
C.~Tang, S.~Chen, L.~Fan, L.~Xu, Y.~Liu, Z.~Tang, and L.~Dou.
\newblock A large-scale empirical study on industrial fake apps.
\newblock In {\em 2019 IEEE/ACM 41st International Conference on Software
  Engineering: Software Engineering in Practice (ICSE-SEIP)}, pages 183--192.
  IEEE, 2019.

\bibitem{dnstwist2}
K.~Tian, S.~T. Jan, H.~Hu, D.~Yao, and G.~Wang.
\newblock Needle in a haystack: Tracking down elite phishing domains in the
  wild.
\newblock In {\em Proceedings of the Internet Measurement Conference 2018},
  pages 429--442, 2018.

\bibitem{wang2018beyond}
H.~Wang, Z.~Liu, J.~Liang, N.~Vallina-Rodriguez, Y.~Guo, L.~Li, J.~Tapiador,
  J.~Cao, and G.~Xu.
\newblock Beyond google play: A large-scale comparative study of chinese
  android app markets.
\newblock In {\em Proceedings of the Internet Measurement Conference 2018},
  pages 293--307, 2018.

\bibitem{wang2006strider}
Y.-M. Wang, D.~Beck, J.~Wang, C.~Verbowski, and B.~Daniels.
\newblock Strider typo-patrol: Discovery and analysis of systematic
  typo-squatting.
\newblock {\em SRUTI}, 6(31-36):2--2, 2006.

\end{thebibliography}

\end{document}